\documentclass[aps,prd,longbibliography,nofootinbib]{revtex4-2}

\usepackage{graphicx}
\usepackage{amsmath,amssymb,bm}
\usepackage{siunitx}
\usepackage{booktabs}
\usepackage{microtype}
\usepackage{natbib}
\usepackage{hyperref}
\usepackage{array}
\usepackage{ragged2e}
\usepackage{xspace}
\usepackage{comment}
\hypersetup{colorlinks=true,allcolors=blue}
\usepackage[percent]{overpic}

\newcommand{\kms}{\mathrm{km\,s^{-1}}}

\newcommand{\yr}{\mathrm{yr}}
\newcommand{\LEO}{LEO\xspace}

\newcommand{\PMD}{PMD\xspace}
\newcommand{\SSA}{SSA\xspace}
\newcommand{\STM}{STM\xspace}
\newcommand{\ADR}{ADR\xspace}
\newcommand{\JCA}{JCA\xspace}
\newcommand{\CAR}{\mathrm{CAR}}
\newcommand{\ERI}{\mathrm{ERI}}
\newcommand{\AOU}{\mathrm{AOU}}
\newcommand{\OYF}{\mathrm{OYF}}
\newcommand{\DMR}{\mathrm{DMR}}

\newcommand{\Hs}{\mathcal{H}}

\newcommand{\Dmg}{\mathcal{D}}

\newcolumntype{L}[1]{>{\raggedright\arraybackslash}p{#1}}
\newcolumntype{C}[1]{>{\centering\arraybackslash}p{#1}}
\newcolumntype{R}[1]{>{\raggedleft\arraybackslash}p{#1}}
\begin{document}

\title{Orbital Debris in Earth Orbit: Operations, Stability, Control, and Market Formation}

\author{Slava G. Turyshev}
\affiliation{
Jet Propulsion Laboratory, California Institute of Technology,\\
4800 Oak Grove Drive, Pasadena, CA 91109-0899, USA
}%

\date{\today}

\begin{abstract}
Orbital debris is a nonlinear control problem in a stratified orbital environment, not a static inventory. This paper develops a reduced-order shell-and-size framework that connects collision-rate scaling, fragment-production gain, natural and controlled sinks, and orbital residence time to intervention ranking and procurement design. The formulation identifies three dominant control levers for near-term orbital sustainability: high-confidence disposal and short post-failure residence time for new spacecraft; reduced encounter-plane covariance for the high-risk conjunction tail; and retirement or deflection of the residual hazard stock of long-lived inactive bodies. A source-gain/sink stability margin separates shells that are operationally crowded but dynamically damped from shells that are dynamically amplifying. The analysis distinguishes the traffic-driven workload peak near 500--600 km from the persistence-driven hazard peak near $\sim$850 km, where inactive mass and long lifetime dominate future fragment production. Current public statistics report $\sim$44,870 tracked objects and more than 16,200 tonnes of material in Earth orbit, with model populations far larger below routine-catalog thresholds. The resulting intervention stack is rapid post-mission disposal, targeted covariance improvement for high-risk encounters,  selective just-in-time collision avoidance or active removal of high-hazard derelicts. The appropriate procurement metric is not the number of objects removed, but verified reduction in time-integrated environmental hazard: verified disposal, verified reduction in ambiguous high-risk conjunctions,  verified reduction in residual hazard stock.

\end{abstract}

\maketitle


\section{Introduction}
\label{sec:intro}

Orbital debris is no longer well represented as a by-product of space activity; it is a control problem on the orbital environment. Two features make the problem technically distinctive. First, for a populated orbital shell the expected collision rate scales approximately with the square of local occupancy, so traffic growth produces a nonlinear increase in conjunction burden and residual collision exposure. Second, the consequence of failure is discontinuous: a single catastrophic collision can generate a long-lived fragment cloud that raises hazard for many other objects over years to centuries. The resulting system is path-dependent. Once a shell accumulates sufficient inactive mass, the environment may continue to deteriorate even if later launch traffic stabilizes or even stops \cite{ESA2025SER,ESA2025EnvWeb,NASAOTPS2023}.

The sharp question addressed here is not whether debris is undesirable, nor whether active removal is beneficial in principle. It is whether the present \LEO environment is entering a regime in which operational burden and long-horizon hazard grow faster than improved mitigation compliance can suppress them, and, if so, which control variables dominate the system response. In this paper those variables are: (1) disposal reliability $p_{\rm disp}$ for newly launched spacecraft; (2) encounter-state uncertainty through the combined encounter-plane covariance $C_b$ for the high-risk conjunction tail; and (3) residual hazard stock $\Pi_k$ for inactive high-mass legacy objects. The central thesis is that orbital sustainability over the next decade will be determined primarily by how aggressively these three variables are reduced in the specific shells and object classes that dominate
long-horizon instability.

This paper develops a reduced-order control framework for orbital-debris intervention ranking. It does not attempt to replace calibrated evolutionary environment models. Instead, it connects current observables, shell-level hazard measures, operator workload, and intervention-specific decision variables within a single comparative architecture. Four evidentiary classes are kept distinct throughout: catalogued or measured quantities, model-based population estimates, long-horizon scenario outputs, and scenario-conditioned economic results. Those categories do not carry the same evidentiary status and should not be interpreted as interchangeable.

This paper makes five technical contributions. First, it reframes orbital debris as a coupled operations--stability problem rather than a static inventory problem. Second, it introduces a reduced-order control formulation in which near-term sustainability is governed primarily by disposal reliability $p_{\rm disp}$, encounter-state uncertainty through $C_b$, and legacy hazard stock $\Pi_k$. Third, it connects those controls to current operator experience through constellation-scale maneuver burden. Fourth, it maps the intervention space into explicit decision rules for post-mission disposal (\PMD), just-in-time collision avoidance (\JCA), and active debris removal (\ADR) under cost and uncertainty. Fifth, it identifies the procurement, verification, liability, and benefit-incidence conditions under which debris-control services can be sustained. Beyond technical interpretation, the framework developed here is intended to support contract design, public procurement, and allocation of responsibility for verified risk reduction.

The framework is also consistent with the recent IAF--IAA--IISL Technical Committee 26 work on space traffic management. That literature treats \STM as a coupled technical, operational, legal, and coordination problem rather than as a narrow conjunction-screening problem \cite{Bonnal2025STM}. The most relevant components for the present paper are the TC26 treatments of large constellations \cite{Sorge2025LargeConstellations}, collision avoidance and risk mitigation \cite{Sorge2025COLA}, orbital-data precision and covariance realism \cite{DoladoPerez2025OrbitalPrecision}, shared catalogues and data-fusion architectures \cite{Kerr2025SharedCatalog,Oltrogge2025DataFusion}, and the persistent gap between debris-mitigation rules and effective compliance \cite{Spencer2025Compliance}. These studies support the paper's treatment of debris control as a coupled problem involving tracking quality, operator coordination, mitigation performance, legal authority, and enforceable incentives.

The novelty of the present paper does not lie in restating that debris is harmful or that remediation can be beneficial in principle. It lies in showing that a common reduced-order structure links three control variables---$p_{\rm disp}$, $C_b$, and $\Pi_k$---to shell diagnosis, intervention ranking, procurement variables, and contractible performance outputs.

We address the stock--flow dynamics of the debris environment in \LEO and the interventions that change its long-horizon trajectory. Launch-phase collision-on-launch-avoidance constraints for crewed or high-value missions are operationally important symptoms of orbital crowding, but they concern short-duration ascent safety rather than the state variables analyzed here. They are therefore treated as an operational boundary condition rather than as a core element of long-horizon debris control.

The paper is organized to move from operations and environment, to orbital dynamics and control, and then to implementation and financing. Section~\ref{sec:acute} shows why the problem is acute now, emphasizing constellation-scale maneuver burden and present environment observables. Section~\ref{sec:trends} diagnoses recent traffic, fragmentation, disposal, and long-horizon stability trends. Section~\ref{sec:physics} develops the shell-level hazard proxy, a minimal shell-and-size population balance, and a local stability criterion. Section~\ref{sec:interventions} evaluates the intervention physics and operational decision rules for rapid disposal, improved state knowledge, \ADR/\JCA, centimeter-class remediation, and shielding. Section~\ref{sec:implementation} describes NASA, NOAA, FCC, ESA, and industry implementation roles. 
Section~\ref{sec:econ} then treats debris as a procurement and mechanism-design problem, including portfolio optimization, payment design, and allocation of responsibility.
Section~\ref{sec:program} translates these results into a priority program for 2026--2035, and Section~\ref{sec:concl} concludes.

\section{Why the problem is acute now: operational burden and present environment}
\label{sec:acute}

\subsection{Constellation-scale collision-avoidance burden}
\label{sec:starlink_case}

\begin{figure}[t]
\includegraphics[width=0.70\linewidth]{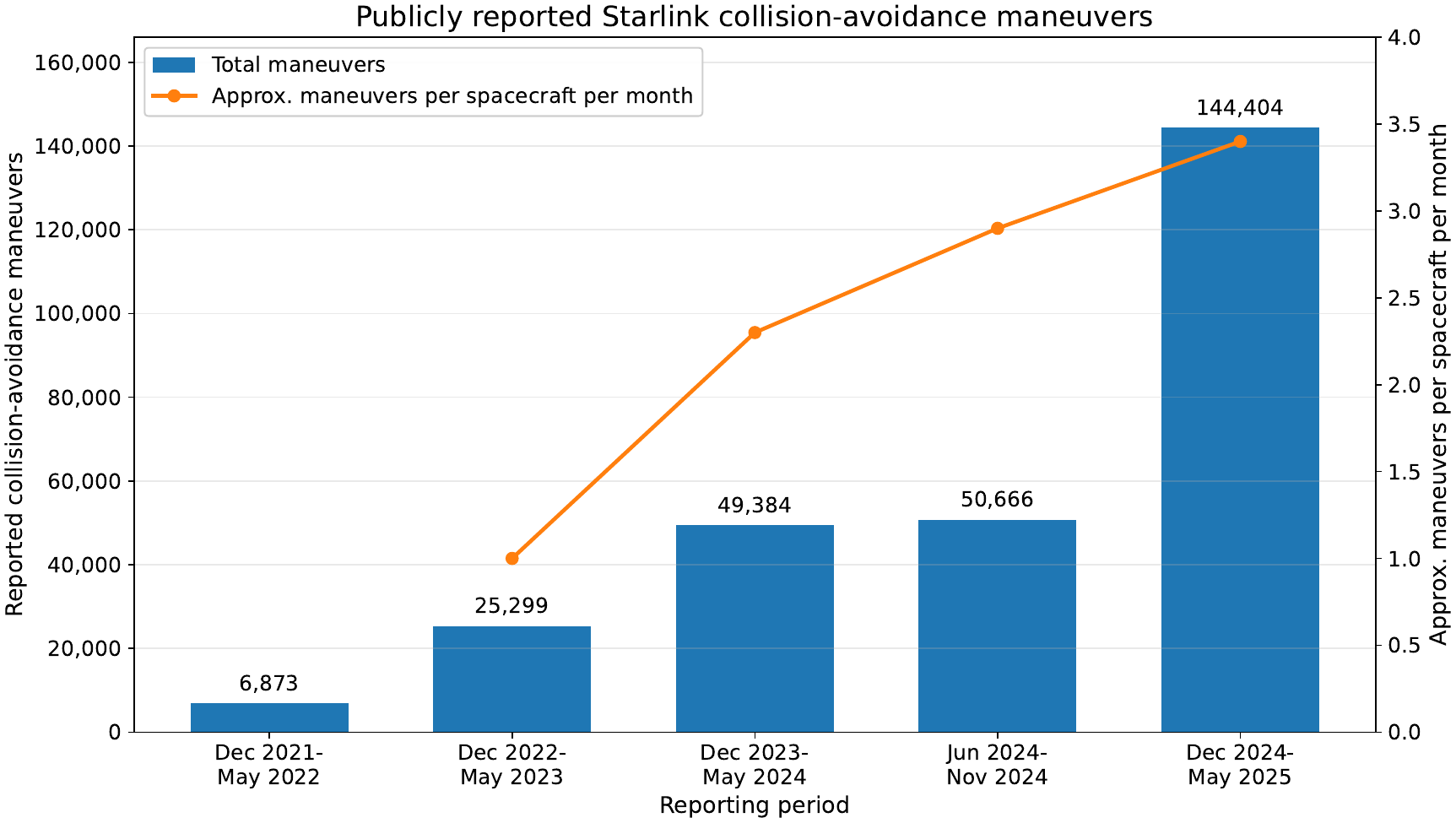}
\caption{Starlink collision-avoidance maneuver  workload by contiguous 6-month reporting period. Blue bars show total reported maneuvers in each period. The orange secondary axis shows approximate maneuvers per spacecraft per month where a direct or reconstructable per-spacecraft rate is available from the public filing record or closely linked public analyses of those filings \cite{SpaceXSemiannual2023Jun,SpaceXSemiannual2023Dec,SpaceXSemiannual2024Jul,SpaceXSemiannual2024Dec,SpaceXSemiannual2025Jul,StarlinkAerospaceAmerica2025}. The earliest interval (Dec 2021--May 2022) is retained to show operational scale, but no directly comparable primary-source per-spacecraft rate is plotted for that interval.}
\label{fig:starlink}
\end{figure}

The debris problem is no longer visible only in static population counts; it is already visible in industrial-scale flight operations. The clearest public case is Starlink. SpaceX reported 6,873 collision-avoidance maneuvers in the six months from Dec 2021 through May 2022, with more than 1,700 of those maneuvers driven by debris from the Nov 2021 Cosmos~1408 anti-satellite test \cite{SpaceXStarlink2022,Pardini2023Cosmos1408}. Public reporting based on later FCC-filed SpaceX semiannual reports indicates a rapid rise to 13,612 maneuvers in June--Nov 2022, 25,299 in Dec 2022--May 2023, 24,410 in June--Nov 2023, 49,384 in Dec 2023--May 2024, 50,666 in June--Nov 2024, and 144,404 in Dec 2024--May 2025 (Figure~\ref{fig:starlink}) \cite{SpaceXSemiannual2023Jun,SpaceXSemiannual2023Dec,SpaceXSemiannual2024Jul,SpaceXSemiannual2024Dec,SpaceXSemiannual2025Jul,StarlinkAerospaceAmerica2025}. These totals are essential because they communicate the scale of the operational burden. However, the burden cannot be interpreted from raw maneuver counts alone because the Starlink fleet was also growing rapidly over the same period.

Raw maneuver counts are not themselves a complete environment metric. They scale jointly with active fleet size, object density in the occupied shells, covariance realism, and the operator’s chosen maneuver trigger. The maneuver trigger $P_c^\star$ denotes the collision-probability threshold at which a predicted conjunction is escalated from monitoring to active avoidance planning or execution. It is therefore both a policy variable and a safety variable: for fixed shell occupancy and covariance quality, reducing $P_c^\star$ from $\sim 10^{-5}$ toward $\sim 10^{-6}$ will generally increase the number of conjunctions that are acted upon and thus raise the observed maneuver rate.

Public filings and regulatory documentation indicate that SpaceX's
maneuver trigger evolved over time. The public record is most safely interpreted as a threshold-era sequence rather than as a precisely dated time series: early Gen2 reporting used a trigger near $P_c^\star\simeq 10^{-5}$; later reporting and public analyses indicate an intermediate trigger near $5\times 10^{-6}$; and FCC documentation by early 2026 described SpaceX as maneuvering when conjunction risk reached
approximately $10^{-6}$ \cite{SpaceXSemiannual2023Jun,
SpaceXSemiannual2024Dec,SpaceXSemiannual2025Jul,FCC2026SpaceXThreshold}. The exact transition dates are not uniformly resolved in the public record and should not be inferred from raw maneuver counts alone.

Although the regulatory and operator literature usually uses the term ``collision-avoidance maneuver,'' the executed action is more precisely a risk-reduction maneuver. It does not guarantee avoidance of a collision; it reduces the estimated probability of collision for a specific conjunction, usually subject to an operator-specific post-maneuver abatement goal. A useful scalar measure of abatement is
\begin{equation}
\Delta_{\log P_c}
\equiv
\log_{10}\!\left(\frac{P_{c,\rm pre}}{P_{c,\rm post}}\right).
\label{eq:pc_abatement}
\end{equation}
For example, a reduction from $1.4\times 10^{-4}$ to
$3.1\times 10^{-6}$ corresponds to $\Delta_{\log P_c}\simeq 1.45$, approximately the 1.5-order-of-magnitude abatement commonly used in NASA conjunction-mitigation guidance
\cite{NASA_CA_S3VI_2021}. Thus, the operational cost-benefit variable is not simply the number of maneuvers, but the aggregate reduction in high-tail $P_c$ achieved per unit operational cost, propellant use, schedule disruption, and induced secondary-conjunction risk.

A more decision-relevant measure is maneuver intensity per active spacecraft. The public record supports a clear rise in this quantity. Publicly documented or reconstructed rates are approximately 1 maneuver per spacecraft per month in June--Nov 2022, Dec 2022--May 2023, and June--Nov 2023; about 27 maneuvers per spacecraft-year in Dec 2023--May 2024 (about 2.3 per spacecraft per month); about 35 maneuvers per spacecraft-year in June--Nov 2024 (about 2.9 per spacecraft per month); and roughly 37--44 maneuvers per spacecraft-year by generation in Dec 2024--May 2025 (about 3.1--3.7 per spacecraft per month) \cite{SpaceXSemiannual2023Jun,SpaceXSemiannual2023Dec,SpaceXSemiannual2024Jul,SpaceXSemiannual2024Dec,SpaceXSemiannual2025Jul}. Figure~\ref{fig:starlink} therefore reports not only total maneuver counts, but also approximate maneuvers per spacecraft per month where the public record supports that normalization.

A more useful normalized quantity is the collision-avoidance rate (CAR) per active satellite-year at a declared collision-probability threshold, $P_c^\star$, 
\begin{equation}
\CAR(P_c^\star) \equiv \frac{M_{\rm CA}}{N_{\rm sat}\,\Delta t},
\label{eq:car}
\end{equation}
where $M_{\rm CA}$ is the number of executed collision-avoidance maneuvers during interval $\Delta t$ and $N_{\rm sat}$ is the mean active fleet size during that interval. Without normalization by both fleet size and trigger threshold, cross-operator or cross-period comparisons are analytically misleading.
 Eq.~(\ref{eq:car}) is intended to separate environment pressure from operator policy. A rise in $\mathrm{CAR}(P_c^\star)$ at fixed threshold indicates increasing operational burden per active spacecraft, whereas a rise in raw maneuver count alone may also reflect fleet growth or a more conservative maneuver trigger. For that reason, future operator disclosure should always report maneuver counts together with $N_{\rm sat}$ and the adopted $P_c^\star$.

A useful decomposition of maneuver burden is
\begin{equation}
M_{\rm CA} \approx N_{\rm sat}\,\Delta t\,
\Lambda_{\rm esc}\!\left(P_c^\star; n_i,\sqrt{\det C_b},\Theta_{\rm op}\right),
\label{eq:workload_model}
\end{equation}
where $\Lambda_{\rm esc}$ is the per-spacecraft escalation rate into maneuver decisions and $\Theta_{\rm op}$ denotes operator policy variables such as screening horizon, maneuver threshold, autonomy logic, duty constraints, and maneuver deadbands; also 
 $n_i$ denotes a shell-local occupancy proxy and $C_b$ denotes an encounter-plane covariance proxy; both are introduced formally in Sec.~\ref{sec:physics}. At this stage they are used only to emphasize that maneuver burden depends jointly on environment density, state uncertainty, and operator policy. 
In practical terms, Eq.~(\ref{eq:workload_model}) shows that maneuver burden should be interpreted as the product of fleet scale, reporting interval, conjunction-escalation intensity, and operator policy. This is why the same raw maneuver count can imply very different operational states if fleet size or maneuver threshold has changed.

Eq.~(\ref{eq:workload_model}) explains why the threshold-era information in Figure~\ref{fig:starlink} matters: a reduction in the maneuver trigger from $P_c^\star\simeq10^{-5}$ toward $\simeq10^{-6}$ raises maneuver burden even if the physical environment were unchanged, so the figure must be read as a combination of environment pressure, fleet growth, and a more conservative collision-avoidance policy.

\subsection{Population, mass, shell structure, and altitude separation}

\begin{figure}[t]
\includegraphics[width=0.50\linewidth]{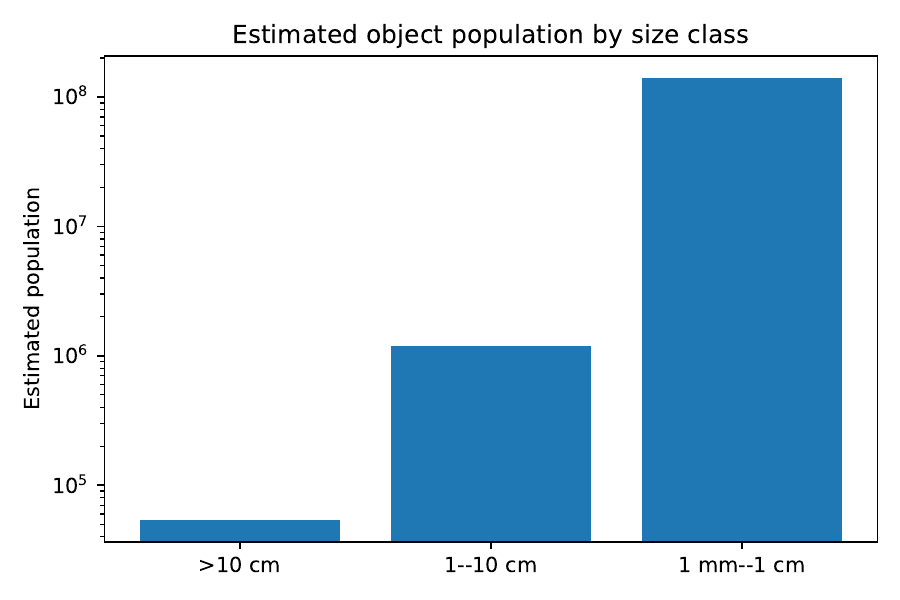}
\caption{Estimated Earth-orbit object population by size regime based on ESA statistics updated on 21 April 2026 \cite{ESA2026Stats}. The 1--10 cm regime is operationally significant because it is much more numerous than the publicly tracked population yet energetic enough to be mission-ending for many spacecraft.}
\label{fig:population}
\end{figure}

ESA's public statistics, updated on 21 April 2026, report about 44,870 regularly tracked objects in Earth orbit, about 17,610 satellites still in space, about 15,200 functioning satellites, and more than 16,200 tonnes of total orbiting mass \cite{ESA2026Stats}. Model-based estimates place the environment at roughly 54,000 objects larger than 10 cm, 1.2 million objects from greater than 1 cm to 10 cm, and 140 million
objects from greater than 1 mm to 1 cm \cite{ESA2026Stats}. The current catalogued LEO population is about 26,400 objects with combined mass of about 9,180 tonnes. Table~\ref{tab:envstate} and Fig.~\ref{fig:population} summarize these values.

\begin{table*}[t]
\caption{Representative current state of the Earth-orbit debris environment from ESA public statistics and the 2025 ESA environment assessment \cite{ESA2026Stats,ESA2025SER}.}
\label{tab:envstate}
\begin{tabular}{l c}
\toprule
{Metric} & {Value} \\
\midrule
Tracked objects in Earth orbit & about 44,870 \\
Satellites still in space & about 17,610 \\
Functioning satellites & about 15,200 \\
Total orbiting mass & $>16,200$ tonnes \\
Catalogued objects in LEO & about 26,400 \\
Mass in LEO & about 9,180 tonnes \\
Estimated population $>10$ cm & 54,000 \\
Estimated population $>1$ cm--10 cm & $1.2\times 10^6$ \\
Estimated population $>1$ mm--1 cm & $1.4\times 10^8$ \\
Newly catalogued fragments added in 2024 & $>3{,}000$ \\
Intact object reentries in 2024 & 1,200 \\
\bottomrule
\end{tabular}
\end{table*}

Population growth is not uniform in altitude. ESA's 2025 report shows a pronounced concentration of active payloads around 500--600 km, while the long-horizon environmental index peaks higher, near $\sim 850$ km at inclinations around 70--80 degrees when 90\% \PMD success for active objects is assumed \cite{ESA2025SER,ESA2025EnvWeb}. The physical reason is straightforward. In the lower peak, local occupancy drives conjunction workload. In the higher peak, weak drag and large inactive mass amplify persistence and breakup consequence. Operational burden is therefore controlled by one shell regime, while long-term instability is controlled by another.

\subsection{Hazard regimes by size and altitude}

The debris environment is best separated into three engineering size regimes, summarized in Table~\ref{tab:sizehazard}. The $>10$ cm class is often catalogued well enough for conjunction screening, although not always with the covariance quality required for efficient operations. The 1--10 cm class is the most operationally difficult because it is usually not maintained in routine custody and can be mission-terminating for many operational spacecraft. It should not, however, be described uniformly as catastrophic for intact derelict targets: whether an impact causes catastrophic fragmentation depends on target mass, projectile mass, impact geometry, material properties, and relative velocity. The 1 mm--1 cm class is much more numerous, effectively untracked in routine operations, and must usually be addressed through shielding and resilient design rather than real-time avoidance \cite{NASA2023CAHandbook,
NASAODPORadar,NASAHVIT}.

\begin{table*}[t]
\caption{Engineering hazard regimes for orbital debris.}
\label{tab:sizehazard}
\begin{tabular}{c l l l}
\toprule
{Size} & {Tracking status} & {Typical consequence} & {Dominant control methods} \\
\midrule
$>$10 cm & Largely catalogued  & Catastrophic collision or spacecraft & Conjunction assessment, maneuvering, co- \\
 & in \LEO; screening  & loss; main parent class for severe  & variance improvement, selective remediation  \\
& feasible & fragmentations &  of high-risk derelicts \\

1--10 cm & Incomplete custody; & Mission-terminating for many   space- & Source suppression, rapid disposal of parent  \\
 & difficult for routine  &  craft;  catastrophic fragmentation & objects, improved sensing, selective removal  \\
 &  avoidance & 
 only for sufficiently energetic upper- & or nudging \\
  &   & subrange impacts or low-mass targets
  &  \\

1 mm--1 cm & Effectively untracked  & Penetration, spall, subsystem dam- & Whipple or stuffed-Whipple shielding, re- \\
&  for routine operations & age, possible mission loss depend-& silient layout, fault tolerance, upstream  \\
&  & ing on local shielding and geometry & environment suppression \\
\bottomrule
\end{tabular}
\end{table*}

NASA's conjunction handbook states that in most conjunctions the relative speed usually exceeds \SI{10}{\kilo\meter\per\second} \cite{NASA2023CAHandbook}. The same handbook defines the catastrophic/non-catastrophic boundary through the specific relative kinetic energy
\begin{equation}
\varepsilon = \frac{\min(M_1,M_2)}{\max(M_1,M_2)}\,\frac{V_{\rm rel}^2}{2},
\label{eq:epsilon}
\end{equation}
with catastrophic breakup expected when $\varepsilon > 4\times 10^4~\mathrm{J\,kg^{-1}}$ \cite{NASA2023CAHandbook}. At \SI{10}{\kms}, the kinetic energy per unit mass is \SI{50}{\mega\joule\per\kilogram}, which explains why centimeter-class impacts are dangerous even when the projectile mass is modest.

Eq.~(\ref{eq:epsilon}) also clarifies why the 1--10 cm bin should be
interpreted as mission-terminating rather than uniformly catastrophic. For a small projectile striking a much larger target, the catastrophic condition can be approximated as
\begin{equation}
\frac{m_p}{M_t}
>
\frac{2\varepsilon_{\rm cat}}{V_{\rm rel}^2}
=
8.0\times 10^{-4}
\left(\frac{10~{\rm km~s^{-1}}}{V_{\rm rel}}\right)^2,
\label{eq:cat_mass_ratio}
\end{equation}
where $\varepsilon_{\rm cat}=4\times 10^4~{\rm J\,kg^{-1}}$.
At $V_{\rm rel}=10~{\rm km~s^{-1}}$, catastrophic breakup of a
$100$ kg target requires a projectile mass of order $80$ g, whereas a $500$ kg target requires a projectile mass of order $0.4$ kg. For an aluminum-density spherical projectile, these correspond to diameters of approximately 3.8 cm and 6.6 cm, respectively. Thus, the lower part of the 1--10 cm population is typically a mission-loss hazard, while the upper part can become an environmental-fragmentation hazard for many robotic spacecraft.

\section{Observed trends and expected evolution} \label{sec:trends} 

\subsection{Traffic growth, fragmentations, and reentries}

The recent environment trends are driven by two distinct processes: rapid growth in active traffic and continued fragmentation of inactive objects. ESA reports that while the exponential growth in payload count slowed in 2024 relative to the immediate preceding surge years, the number of launches continued to rise and total launch mass and area remain at historical highs \cite{ESA2025SER}. At the same time, the number and scale of commercial constellations in certain LEO bands continue to increase year over year \cite{ESA2025EnvWeb}.

Fragmentation remains a chronic source term. ESA reports that over the last two decades an average of 10.5 non-deliberate fragmentation events per year has occurred, with the rate dropping to 1.7 per year when the lifetime of generated fragments is given greater weight \cite{ESA2025SER}. In 2024, several major and several smaller fragmentation events added more than 3,000 newly catalogued fragments \cite{ESA2025SER,ESA2025EnvWeb}. A practical implication is that the environment is not dominated by a smooth, predictable source term; rather, much of the annual increment in trackable debris may arrive in punctuated bursts.

Reentry activity has also increased. ESA reports that 1,200 intact objects re-entered in 2024, and the public summary notes that intact objects are now re-entering the atmosphere on average more than three times per day \cite{ESA2025SER,ESA2025EnvWeb}. This increasing reentry rate is partly a positive consequence of improved disposal behavior and elevated solar activity, but it also means the industry is moving into a regime where reentry-risk management and controlled-disposal design will become increasingly important.

\subsection{Improving compliance, but not enough}

Post-mission disposal performance is improving in aggregate, especially for rocket bodies, but aggregate compliance should not be interpreted as uniform compliance across all launch campaigns or altitude regimes. ESA's 2025 summary states that approximately 90\% of rocket bodies in \LEO now vacate valuable orbits within the legacy 25-year limit, while about 80\% already satisfy ESA's more demanding 5-year standard \cite{ESA2025EnvWeb}. The same source reports that controlled rocket-body reentries outnumbered uncontrolled ones for the first time in 2024 \cite{ESA2025EnvWeb}. These are meaningful improvements, but they do not eliminate the risk created by a comparatively small number of new upper-stage objects inserted into persistent high-altitude \LEO shells. For a newly abandoned rocket body $r$, the incremental hazard-stock contribution scales approximately as
\begin{equation}
\Delta \Pi_r
\propto
m_r\,\chi_r\,\phi(h_r,i_r)\,\tau(h_r,A_r/m_r),
\label{eq:rocket_body_increment}
\end{equation}
where $m_r$ is intact mass, $\chi_r$ is a breakup-severity factor,
$\phi(h_r,i_r)$ is the local collision-flux proxy, and $\tau$ is the
residual residence time. Eq.~(\ref{eq:rocket_body_increment}) explains why a small number of recently abandoned high-altitude upper stages can materially offset broad aggregate improvements in rocket-body disposal: the controlling quantity is not only the count of compliant rocket bodies, but the mass-weighted and lifetime-weighted tail of non-compliant objects in persistent shells.

Recent Chinese constellation deployments also illustrate why aggregate rocket-body compliance should not be read as uniformly low-risk. Public catalogues and sector analyses indicate that some Guowang/Xingwang and Qianfan deployment stages have been left in persistent high-altitude LEO orbits. In the shell-level hazard budget, such objects enter through intact mass, local collision flux, breakup severity, and residual lifetime rather than through object count alone. If a numerical count is reported, it should be tied to a stated catalogue source, query date, object-class definition, altitude criterion, and treatment of fragmented or decayed stages.

However, payload clearance remains less satisfactory in the aggregate, and the absolute environment continues to worsen because current compliance is not sufficient relative to the growth in traffic and the size of the legacy dead-mass inventory. ESA states directly that not enough satellites leave heavily congested orbits at end of life, that mitigation compliance is improving only slowly, and that the improvement is not enough to stop the increase in debris \cite{ESA2025EnvWeb}.

In the United States, the FCC's 5-year post-mission disposal rule is now in force for missions ending in or passing through LEO below 2,000 km and disposing via uncontrolled atmospheric reentry \cite{FCC2025Guide}. That change is significant, but compliance reliability must be judged not only by the rule on paper but by the actual probability of spacecraft self-disposal success in the presence of failures.

\subsection{Long-term environment evolution}

ESA's long-horizon simulations make two points that should shape engineering policy. First, even in a no-further-launches scenario, the amount of space debris in \LEO increases because existing objects continue to fragment and collide faster than debris naturally reenters \cite{ESA2025SER}. Second, under the current extrapolation of launch traffic, explosion rates, and disposal success, the simulated \LEO environment continues to deteriorate and the associated risk is predicted to be about 4 times greater than a chosen orbital-sustainability threshold based on a 2014-traffic world with high-level implementation of the IADC mitigation framework, including 90\% success of 25-year PMD \cite{ESA2025SER}. In other words, even a 90\% PMD world is not a sufficient long-term target if traffic and inactive-object inventories remain large.

NASA's 2023 remediation analysis reaches a compatible conclusion from a different direction. That report assumes a 90\% PMD compliance rate and still finds that stabilizing the debris population requires some level of remediation. It further notes that debris below about 650 km naturally deorbits within 25 years, whereas above that altitude the only way to stabilize or reduce debris counts is active removal \cite{NASAOTPS2023}. In the illustrative NASA model, removing 2 large objects per year helps, but removing 5 large objects per year is the more effective stabilizing case under 90\% PMD assumptions \cite{NASAOTPS2023}. 

The specific number is model-dependent, but the qualitative implication is robust under the cited ESA and NASA scenario assumptions: when legacy inactive mass in persistent shells is already large, even high PMD success for new missions may not by itself stabilize the long-horizon environment. In that regime, mitigation suppresses future source terms, but selective remediation or JCA is required to reduce the inherited high-$\Pi_k$ stock.

\begin{figure}[t]
\includegraphics[width=0.65\linewidth]{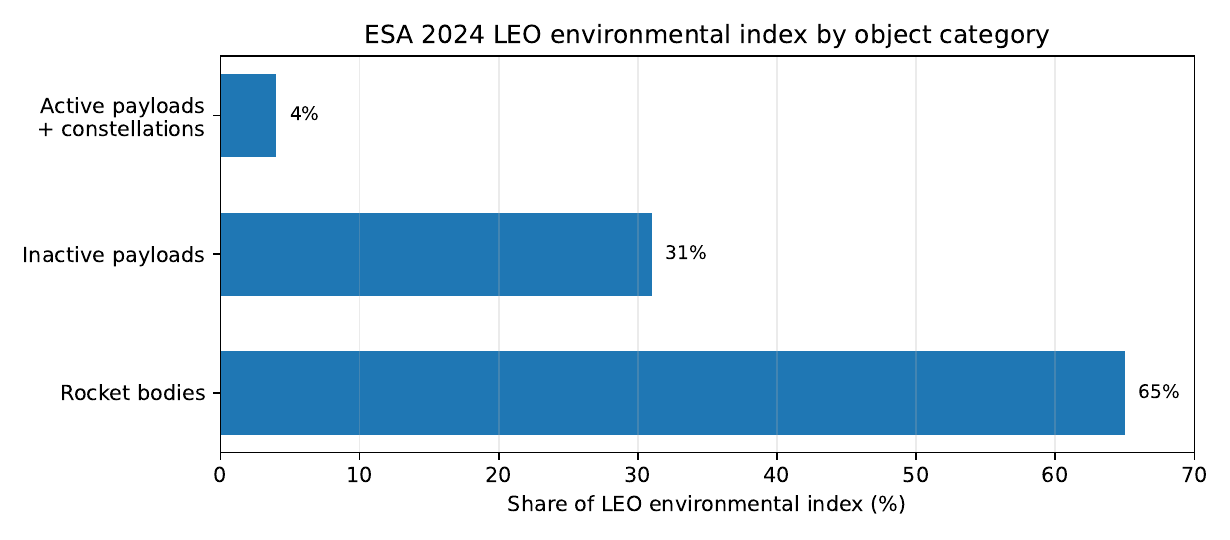}
\caption{Share of ESA's 2024 \LEO environmental index by object category. The index assumes 90\% PMD success for active objects and shows that 96\% of the total risk is associated with inactive objects, dominated by rocket bodies \cite{ESA2025SER}.}
\label{fig:riskshare}
\end{figure}

\subsection{Why the dominant legacy risk is concentrated in inactive objects} 

The strongest indicator for remediation targeting is not object count but environmental index. ESA's 2025 report states that, assuming 90\% PMD success for active objects, 96\% of the total \LEO environmental risk is associated with inactive objects, and that the largest contribution comes from spent rocket bodies \cite{ESA2025SER}. As shown in Figure~\ref{fig:riskshare}, roughly 65\% of the ESA \LEO environmental index is associated with rocket bodies, 31\% with inactive payloads, and only  $\sim$4\% combined with active payloads and constellations. The engineering implication is that the highest-value remediation targets are the relatively small population of large, dead, long-lived objects that dominate future fragment production.

\section{Physics of instability and control}
\label{sec:physics}

\subsection{Shell observables and a hazard proxy}

A compact shell-level description requires four observables: local number density $n_i$, inactive intact mass $M_i^{\rm dead}$, mean collision kernel $\langle \sigma v \rangle_i$, and post-failure residence time $\tau_i$. A useful first-order proxy for the contribution of shell $i$ to long-horizon environmental hazard is
\begin{equation}
\Hs_i \propto n_i^2 \,\langle \sigma v \rangle_i\,\bar S_i\,\tau_i,
\label{eq:hazardproxy}
\end{equation}
where $\bar S_i$ is a representative breakup-severity factor. Eq.~(\ref{eq:hazardproxy}) separates the two altitude regimes: The lower-altitude peak near 500--600 km is dominated by large local occupancy and therefore drives conjunction workload. The higher-altitude peak near $\sim 850$ km is dominated by long residence time and large inactive mass and therefore drives long-horizon instability. A shell can therefore be operationally burdensome without being the principal driver of long-run environmental deterioration, and conversely a shell can dominate long-horizon hazard without dominating present-day maneuver demand.

Eq.~(\ref{eq:hazardproxy}) is most useful when it is used to compare shells explicitly. Consider a lower shell $L$ and higher shell $H$. If
\[
\frac{n_H}{n_L}=0.5,\qquad
\frac{\langle\sigma v\rangle_H}{\langle\sigma v\rangle_L}\simeq 1,\qquad
\frac{\bar S_H}{\bar S_L}=2,\qquad
\frac{\tau_H}{\tau_L}=20,
\]
then
\[
\frac{\Hs_H}{\Hs_L}
\sim
\Big(\frac{n_H}{n_L}\Big)^2
\frac{\langle\sigma v\rangle_H}{\langle\sigma v\rangle_L}
\frac{\bar S_H}{\bar S_L}
\frac{\tau_H}{\tau_L}
\approx
0.25\times 1\times 2\times 20
\approx 10.
\]
Thus a shell with only half the local occupancy can still dominate the long-horizon hazard budget if persistence and breakup consequence are much larger. This is the formal reason the shell that dominates present maneuver workload need not be the shell that dominates long-run environmental instability.

A useful first-order collision-rate expression for shell $i$ is
\begin{equation}
\dot N_{c,i} \approx \frac{1}{2V_i}\sum_{j,k} N_{ij}N_{ik}\,\langle \sigma_{jk} v_{jk}\rangle,
\label{eq:shellcollision}
\end{equation}
where $V_i$ is shell volume, $N_{ij}$ is the object count of category $j$ in shell $i$, and $\langle \sigma_{jk} v_{jk}\rangle$ is the mean collision kernel over collision cross-section and relative speed. For a thin shell, $V_i\approx 4\pi a_i^2\Delta h$, where $a_i$ is the representative shell radius (or semimajor axis for near-circular shells) and $\Delta h$ is the shell thickness. Eq.~(\ref{eq:shellcollision}) makes the near-$N^2$ scaling of conjunction burden explicit.

\subsection{Population-balance dynamics and a local stability criterion}

Let $N_{is}(t)$ denote the object population in altitude shell $i$ and size class $s$. A minimal population-balance model is
\begin{equation}
\dot N_{is}
=
L_{is}(t)
+
B^{\rm frag}_{is}(\mathbf N)
+
B^{\rm coll}_{is}(\mathbf N)
-
\frac{N_{is}}{\tau_{is}}
-
u^{\rm PMD}_{is} N_{is}
-
u^{\rm ADR}_{is} N_{is},
\label{eq:population_balance}
\end{equation}
where $L_{is}$ is the launch/release source term, $B^{\rm frag}_{is}$ and $B^{\rm coll}_{is}$ are fragment-production terms from explosions and collisions, $\tau_{is}$ is natural orbital residence time, and $u^{\rm PMD}_{is}$ and $u^{\rm ADR}_{is}$ are effective controlled-removal rates. A convenient collision source term is
\begin{equation}
B^{\rm coll}_{is}
=
\sum_{j,\ell}
Y_{j\ell\rightarrow s}\,
P^{\rm cat}_{ij\ell}\,
\frac{N_{ij}N_{i\ell}}{2V_i}\,
\langle \sigma v\rangle_{ij\ell},
\label{eq:collision_source}
\end{equation}
where $Y_{j\ell\rightarrow s}$ is mean fragment yield into class $s$ and $P^{\rm cat}_{ij\ell}$ is the catastrophic-breakup probability.

Linearizing about a reference environment-control pair $(\mathbf N^\ast,\mathbf u^\ast)$ gives
\begin{equation}
\delta\dot{\mathbf N}
=
A\,\delta\mathbf N + B\,\delta\mathbf u,
\qquad
A\equiv
\left.\frac{\partial \mathbf f}{\partial \mathbf N}\right|_{\mathbf N^\ast,\mathbf u^\ast},
\qquad
B\equiv
\left.\frac{\partial \mathbf f}{\partial \mathbf u}\right|_{\mathbf N^\ast,\mathbf u^\ast},
\label{eq:linearized_system}
\end{equation}
where $\delta\mathbf N$ and $\delta\mathbf u$ are perturbations about the reference state and control. The environment is locally unstable when the dominant eigenvalue of the state Jacobian satisfies
\begin{equation}
\max \Re[\lambda(A)] > 0,
\label{eq:instability}
\end{equation}
where $\lambda(A)$ denotes the eigenvalue spectrum of $A$. Eq.~(\ref{eq:instability}) is the formal version of the common statement that some shells can continue to degrade even if launch traffic is reduced: in those regimes, collision-generated source terms dominate natural decay and controlled sinks. This is consistent with ESA's no-further-launches results and NASA's long-horizon remediation analyses \cite{ESA2025SER,NASAOTPS2023}.

Eqs~(\ref{eq:population_balance})--(\ref{eq:instability}) imply a practical stability condition for protected shells: the combined sink from natural decay and controlled removal must exceed the linearized gain associated with collision-generated fragment production. In low-persistence shells this condition may be satisfied largely through rapid \PMD and natural drag. In high-persistence shells, however, the same condition typically fails once inactive intact mass becomes sufficiently large, because fragment-production feedback grows faster than passive clearance. This is the formal reason mitigation alone ceases to be sufficient in some shells even when compliance improves.

For practical shell ranking it is useful to separate the source-gain
part of the Jacobian from the sink part. Let the shell-restricted
linearized dynamics be written as
\begin{equation}
\delta\dot{\mathbf N}_i
=
A_i\,\delta\mathbf N_i,
\qquad
A_i
=
J_i^{\rm src}-S_i,
\label{eq:shell_jacobian_split}
\end{equation}
where
\begin{equation}
J_i^{\rm src}
\equiv
\left.
\frac{\partial\left(\mathbf B_i^{\rm frag}
+\mathbf B_i^{\rm coll}\right)}
{\partial \mathbf N_i}
\right|_{\mathbf N^\ast,\mathbf u^\ast},
\qquad
S_i
\equiv
{\rm diag}\!\left(
\tau_{is}^{-1}+u_{is}^{\rm PMD}+u_{is}^{\rm ADR}
\right).
\label{eq:source_sink_split}
\end{equation}
The full local spectral margin is then
\begin{equation}
\mu_i
\equiv
-\alpha(A_i),
\qquad
\alpha(A_i)\equiv \max \Re[\lambda(A_i)].
\label{eq:stability_margin}
\end{equation}
Shells with $\mu_i>0$ are locally damped, whereas shells with
$\mu_i<0$ are locally amplifying. In the special case where the
source-gain and sink matrices are approximately diagonal or weakly non-normal, Eq.~(\ref{eq:stability_margin}) reduces to the intuitive screening estimate
\begin{equation}
\mu_i
\approx
\lambda_{\min}(S_i)-\alpha(J_i^{\rm src}).
\label{eq:stability_margin_approx}
\end{equation}
This form avoids double counting: the sink appears either inside the full Jacobian $A_i$ or explicitly in $S_i$, but not both.  Eq.~(\ref{eq:stability_margin}) is useful because it compresses a high-dimensional evolutionary model into a scalar engineering screening metric for identifying where mitigation, tracking, or remediation will have the largest marginal effect.

Eq.~(\ref{eq:stability_margin}) is most useful as a triage device rather than as a forecast. Its role is to rank shells by the sign and magnitude of the competition between effective sink strength and linearized fragment-production gain. In practice, this means that the equation can guide where to intensify \PMD requirements, where to prioritize high-fidelity tracking, and where to direct public remediation procurement.

\subsection{Illustrative shell ranking from the reduced stability margin}
\label{sec:stability_ranking}

\begin{figure}[t]
\includegraphics[width=0.58\linewidth]{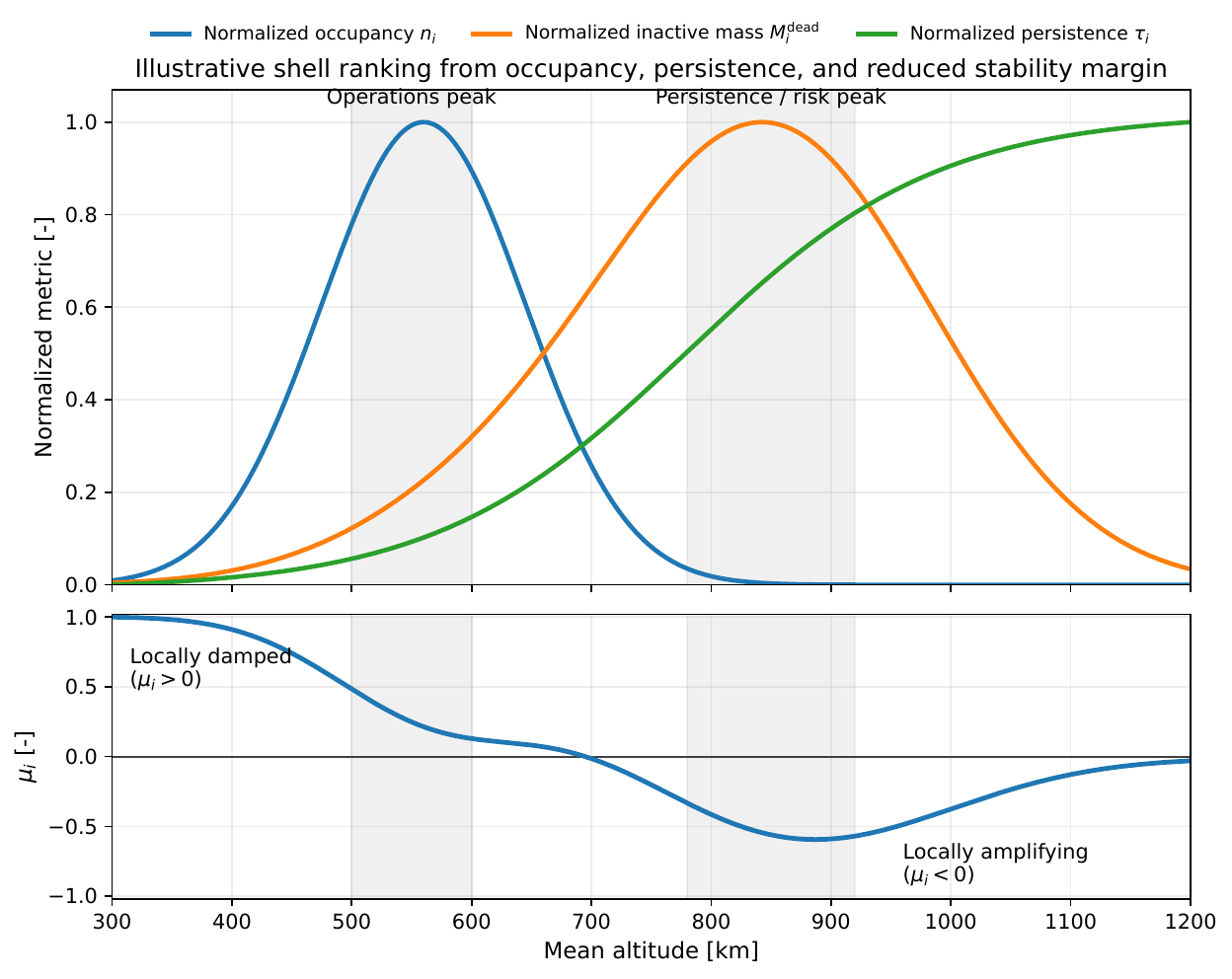}
\vskip -5pt
\caption{Illustrative shell-level ranking using normalized occupancy $n_i$, persistence $\tau_i$, and reduced stability margin $\mu_i$ from Eq.~(\ref{eq:stability_margin}). The figure is intended as a screening diagnostic rather than a full calibrated environment forecast. It separates shells that dominate present conjunction workload from those that dominate long-horizon instability.}
\label{fig:stability_margin}
\end{figure}

Eq.~(\ref{eq:stability_margin}) is useful only if it can distinguish shells that are operationally busy from shells that are dynamically dangerous. Figure~\ref{fig:stability_margin} provides that interpretation by plotting normalized shell occupancy, persistence, and reduced stability margin on a common altitude axis. The key point is that the shell driving current maneuver burden need not be the shell most responsible for long-horizon instability. In the present \LEO environment, the lower constellation band near 500--600~km is expected to dominate conjunction workload because of local occupancy, whereas the band near $\sim850$~km is expected to dominate long-horizon hazard because of long residence time and large inactive intact mass.

The practical decision value of Eq.~(\ref{eq:stability_margin}) is thus as a shell-screening metric. A shell with $\mu_i<0$ is not merely crowded; it is a candidate for intensified mitigation, targeted tracking, or selective remediation. A shell with $\mu_i>0$ may still be operationally busy, but its natural and controlled sinks remain stronger than its linearized fragment-production gain.

\subsection{The three control variables}

The practical implication of Eqs.~(\ref{eq:hazardproxy})--(\ref{eq:instability}) is that the debris-control problem is governed, to first order, by three variables. The first is disposal reliability $p_{\rm disp}$ for newly launched spacecraft. If a spacecraft carries $N_m$ disposal modes with independent success probabilities $p_j$, the effective disposal reliability is
\begin{equation}
p_{\rm disp}=1-\prod_{j=1}^{N_m}(1-p_j).
\label{eq:pdisp}
\end{equation}
Eq.~(\ref{eq:pdisp}) assumes statistical independence among disposal modes. In practice, common-cause failures in power, attitude control, command, propulsion, or thermal survival can dominate the true residual risk, so the engineering objective is not merely to add disposal modes, but to add genuinely independent disposal paths.

The independence assumption in Eq.~(\ref{eq:pdisp}) is optimistic when the disposal modes share power, attitude control, thermal survival, commanding, or propulsion dependencies. A simple common-cause extension is
\begin{equation}
p_{\rm disp}
=
1-
\left[
q_c+(1-q_c)\prod_{j=1}^{N_m}(1-p_j)
\right],
\label{eq:pdisp_common_cause}
\end{equation}
where $q_c$ is the probability of a common-cause failure that disables
all disposal modes. The effect can be large. Two nominally redundant
disposal modes with $p_1=p_2=0.9$ give $p_{\rm disp}=0.99$ under
independence, but with only $q_c=0.05$,
\begin{equation}
p_{\rm disp}
=
1-\left[0.05+0.95(0.1)(0.1)\right]
=
0.9405.
\end{equation}
Thus, redundancy improves environmental performance only when it reduces
the common-cause tail of disposal failure. For licensing and
procurement, the relevant reliability metric is not the number of
disposal modes, but the verified reduction in
$(1-p_{\rm disp})\tau_k$.

The second is encounter-state uncertainty in the high-risk conjunction tail. For a linearized conjunction, the encounter-plane covariance is
\begin{equation}
C_b
=
G\,C_{\rm rel}\,G^T,
\qquad
C_{\rm rel}
=
C_1+C_2-C_{12}-C_{21},
\label{eq:combined_covariance}
\end{equation}
where $C_1$ and $C_2$ are the inertial state covariances of the two objects, $C_{12}$ and $C_{21}$ are cross-covariance terms, and $G$ maps relative-state uncertainty into the encounter plane. If the two state estimates are statistically independent, $C_{12}=C_{21}=0$, and Eq.~(\ref{eq:combined_covariance}) reduces to the common expression
$C_b=G(C_1+C_2)G^T$.

The characteristic ambiguous area in the encounter plane scales with $\sqrt{\det C_b}$, which is why covariance improvement can reduce operator workload. However, the single-event collision probability is not simply proportional to this area. In the short-encounter, Gaussian-error, small-hard-body limit,
\begin{equation}
P_c
\approx
\frac{R_{\rm HBR}^2}{2\sqrt{\det C_b}}
\exp\!\left[
-\frac{1}{2}\mathbf b^T C_b^{-1}\mathbf b
\right],
\label{eq:pc_local_gaussian}
\end{equation}
where $\mathbf b$ is the projected miss vector and $R_{\rm HBR}$ is the hard-body radius. Thus, covariance reduction has two distinct effects: it reduces the ambiguous review area for the high-risk tail, but it can increase or decrease the estimated $P_c$ for a particular encounter depending on the normalized miss distance. The economically relevant quantity is therefore the reduction in high-tail ambiguous conjunctions and risk-reduction decisions, not a universal monotonic decrease in every computed $P_c$.

The operational implication of Eq.~(\ref{eq:combined_covariance}) is concentrated rather than diffuse. Because false-warning burden scales to first order with $\sqrt{\det C_b}$, a tenfold reduction in the characteristic area of the high-risk uncertainty ellipse corresponds, to first order, to a tenfold reduction in ambiguous encounter area for that tail population. The point is not exact proportionality---screening horizon, decision threshold, and autonomy logic still matter---but concentration of value: the first economically important increment of tracking improvement lies in the small subset of conjunctions that dominate manual review, maneuver planning, and conservative action, not in weak global refinement of the full catalogue.

The third is the residual hazard stock of legacy object $k$,
\begin{equation}
\Pi_k = \int_{t_{\rm now}}^{t_{{\rm re},k}}\lambda_k(t)\,S_k(t)\,dt,
\label{eq:Pi_stock}
\end{equation}
where $t_{\rm now}$ is the present decision time, $t_{{\rm re},k}$ is the time of controlled removal, controlled disposal, or natural reentry of object $k$, $\lambda_k(t)$ is the post-failure hazard rate, and $S_k(t)$ is event severity.

Eq.~(\ref{eq:Pi_stock}) explains why high-mass inactive objects in persistent shells dominate remediation priority even when they are not numerically dominant. The integral weights present hazard by remaining residence time, so large intact objects with long lifetime and high breakup severity accumulate large hazard stock even if their instantaneous collision probability is modest. The quantity $\Pi_k$ is therefore the natural bridge between physical risk ranking and economically efficient target selection.

\subsection{What the reduced-order framework implies}
\label{sec:framework_use}

The reduced-order framework is useful only if it changes decisions. Its purpose is not to replace a calibrated environment model, but to identify which control variable is dominant in a given regime and what intervention follows.

At the operational layer, Eqs.~(\ref{eq:car}) and (\ref{eq:workload_model}) show why raw maneuver counts are mixed observables. A rise in $\CAR(P_c^\star)$ at fixed threshold indicates increasing burden per active spacecraft; a rise in raw count alone may instead reflect fleet growth or a more conservative trigger.

At the shell layer, Eqs.~(\ref{eq:hazardproxy})--(\ref{eq:stability_margin}) separate operational crowding from dynamical amplification. A shell with large $n_i$ but $\mu_i>0$ can be operationally busy yet environmentally damped. A shell with more modest instantaneous traffic but $\mu_i<0$ is dynamically dangerous because fragment-production gain exceeds effective sink strength.

At the intervention layer, Eqs.~(\ref{eq:pdisp}), (\ref{eq:combined_covariance}), and (\ref{eq:Pi_stock}) identify the three quantities that actually change the near-term trajectory: $(1-p_{\rm disp})\tau_k$ for new missions, $\sqrt{\det C_b}$ for the high-risk conjunction tail, and $\Pi_k$ for the legacy stock. Eq.~(\ref{eq:budget_selection}) then turns those quantities into a finite-budget portfolio problem, Eq.~(\ref{eq:risk_fee}) into an ex ante pricing signal, Eq.~(\ref{eq:npv_service}) into an economics feasibility condition, and Eq.~(\ref{eq:payment_formula}) into a contract form. The framework is therefore sequential: it begins with observables, moves to shell diagnosis, then to intervention ranking, and finally to procurement and market design.

The reduced-order quantities used here should be interpreted as
screening and decision variables, not as replacements for calibrated environment-evolution models. In particular, $H_i$, $\mu_i$, and $\Pi_k$ are useful for ranking shells, interventions, and target classes, but their absolute values require calibration against higher-fidelity population models and observation-informed catalogues. The framework therefore makes comparative claims: it identifies which physical variables dominate marginal intervention value under specified scenario
assumptions. It does not claim to forecast the future debris population independently of LEGEND-, MASTER-, ORDEM-, or DISCOS-class modeling.

\section{Intervention physics and operational decision rules}
\label{sec:interventions}

\subsection{Rapid post-mission disposal as a reliability control}

The highest-return intervention remains preventing new debris and minimizing the residence time of dead spacecraft. ESA's Zero Debris approach requires no intentional release of mission-related debris, minimization of breakup risk through passivation, self-disposal probability above 90\%, interfaces that permit later removal if self-disposal fails, and \LEO clearance in less than 5 years with cumulative collision probability from launch through disposal below $10^{-3}$ for debris larger than 1 cm \cite{ESAZeroDebris}. The FCC's 5-year disposal rule is aligned with the same direction of travel \cite{FCC2025Guide}.

NASA's 2024 cost-benefit analysis strongly supports rapid disposal. Changing from a 25-year rule to a 15-year rule yields estimated benefits 20--750 times costs and up to about \$6 billion in net benefits over 30 years; a 0-year rule can approach \$9 billion in net benefits under the study assumptions \cite{NASAOTPS2024,Colvin2024IAC}. The same work finds that deorbiting within 5 years or fewer is favored by benefits several hundred times larger than costs, and that drag-device implementations can be even more favorable in optimistic cases, albeit with caveats about increased collision cross-section \cite{Colvin2024IAC}.

From an engineering perspective, disposal should be treated as a reliability problem rather than only a trajectory problem. The expected residual environmental contribution, where the expectation is taken over disposal outcomes and relevant failure states, scales approximately as
\begin{equation}
\mathbb{E}[\Pi_k] \propto (1-p_{\rm disp})\,\tau_k\,\lambda_k\,S_k.
\label{eq:residualhazard}
\end{equation}
Eq.~(\ref{eq:residualhazard}) turns the disposal question into an analytical tool. Relative to a baseline design with $p_{\rm disp}=0.90$ and $\tau_k=25~\yr$, an improved design with $p_{\rm disp}=0.99$ and $\tau_k=5~\yr$ yields
\[
\frac{\mathbb{E}[\Pi_k]_{\rm improved}}{\mathbb{E}[\Pi_k]_{\rm base}}
=
\frac{(1-0.99)\times 5}{(1-0.90)\times 25}
=0.02.
\]
The expected residual contribution falls by a factor of 50. That is why disposal should be treated as a primary system-design variable rather than as an end-of-life compliance afterthought.

The engineering implication is that redundancy matters only to the extent that it materially reduces the product of post-failure residence time and breakup risk in the protected shell. This favors redundant deorbit modes, disposal propellant margin, fail-safe autonomous end-of-life logic, and passive augmentations such as drag devices. NASA's small-spacecraft state-of-the-art review notes, for example, that the DragNET system is a 2.8 kg module with a 14 m$^2$ sail capable of deorbiting a 180 kg spacecraft from 850 km in less than 10 years \cite{NASASmallsatDeorbit}. ESA's ADEO demonstration deployed a 3.6 m$^2$ drag sail from a 10$\times$10$\times$10 cm package in 2022, providing direct in-orbit proof of passive drag augmentation \cite{ESAADEO}.

\subsection{Disposal time as a control knob}
\label{sec:disposal_knob}

Eq.~(\ref{eq:residualhazard}) implies that post-mission disposal is not simply a compliance variable; it is an environmental control knob. At fixed breakup severity and post-failure hazard rate, the expected residual contribution of an object scales approximately with $(1-p_{\rm disp})\tau_k$. This means that disposal policy acts on two margins at once: it can increase the probability of successful removal and reduce the time during which a failed spacecraft remains hazardous.

\begin{figure}[t]
\includegraphics[width=0.62\linewidth]{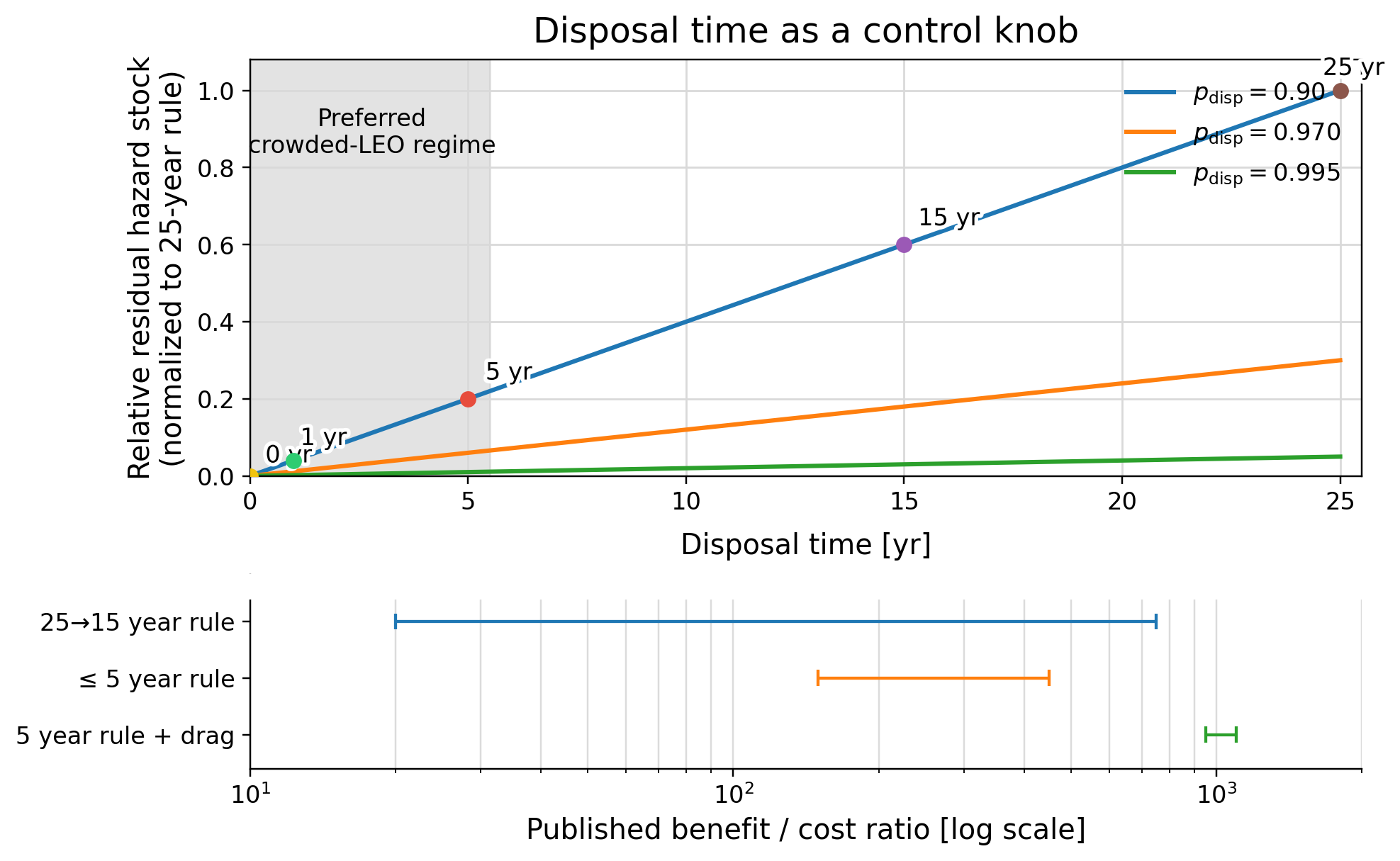}
\caption{Illustrative relation between disposal time, expected residual hazard stock, and published economic value of faster \PMD. The upper panel visualizes the control intuition behind Eq.~(\ref{eq:residualhazard}); the lower panel summarizes published NASA benefit--cost evidence for shorter disposal timelines on a logarithmic axis \cite{NASAOTPS2024,Colvin2024IAC}. The 0-year rule is not shown as a benefit--cost ratio because the published result is expressed primarily as a net benefit of approximately \$9~billion over the study horizon.}
\label{fig:pmd_trade}
\end{figure}

Figure~\ref{fig:pmd_trade} summarizes the intuition behind this result by combining the qualitative scaling from Eq.~(\ref{eq:residualhazard}) with the published NASA benefit--cost estimates for shorter disposal timelines. The figure makes clear why disposal in less than 5~years is not simply a more demanding standard; it is a transition into a substantially lower environmental-liability regime.

\subsection{Tracking, covariance quality, and traffic coordination}

Improved \SSA/\STM is the second pillar of an effective debris strategy. Better state knowledge does not remove debris, but it reduces both missed warnings and false positives. NASA's Orbital Debris Program Office uses radar, optical systems, returned-surface analysis, and laboratory testing to characterize debris populations \cite{NASAODPORadar,NASAODPOMain}. HUSIR is NASA's principal statistical radar source for debris from a few millimeters to the public-catalog threshold, and Goldstone extends sensitivity to millimeter-class metallic spheres under favorable conditions \cite{NASAODPORadar}. On the European side, ESA's Iza\~na-1 station uses short laser pulses to determine distance, velocity, and orbit of space objects with millimeter precision, explicitly to reduce false alarms and unnecessary maneuvers \cite{ESAIzana1}.

NASA's conjunction-assessment handbook emphasizes process discipline as well as sensing: ephemerides with covariances should be furnished routinely, maneuver intentions should be shared in advance, and standard data products such as OEM ephemerides and maneuver reports should be exchanged so that the conjunction-assessment ecosystem has better inputs \cite{NASA2023CAHandbook}. These are prerequisites for scaling constellation operations without driving maneuver rates or missed-risk rates to unacceptable values.

The economics of better state knowledge are unusually strong. NASA's 2024 study concludes that on-demand tracking or uncertainty reduction for high-risk conjunctions has robustly positive net benefits; in optimistic cases benefits exceed costs by more than 100 times. The study also estimates that reducing uncertainty about high-risk conjunctions by a factor of 10 may be worth about \$1.5 billion over 30 years, with diminishing returns beyond the first order-of-magnitude reduction \cite{NASAOTPS2024,Colvin2024IAC}. This is why order-of-magnitude covariance reduction for a small subset of high-risk conjunctions can have a larger operational payoff than modest improvements distributed across the entire catalogue.

\subsection{ADR and JCA as hazard-stock controls}

Active debris removal should be applied selectively, not indiscriminately. The best targets are the highest-risk inactive payloads and rocket bodies in persistent crowded shells. NASA's 2023 remediation analysis evaluates the ``Top 50'' statistically most concerning derelicts and finds that the total benefits of remediating them rise to about \$1.1 billion over 25 years in the illustrative case \cite{NASAOTPS2023}. The 2024 NASA phase-2 study shows that removing 50 large pieces of debris has benefit--cost ratios spanning roughly 0.1 to 10 over 30 years, depending on assumptions about cost and environment evolution \cite{NASAOTPS2024}. Those wide ranges mean that large-object removal is not universally dominant at every cost point, but it can be highly favorable when applied to the right objects.

A major insight from the NASA work is that the most effective remediation is not always permanent removal. \JCA for large derelicts---nudging them only when a threatening conjunction is predicted---can be more cost-effective than full removal in some cases. NASA's 2024 IAC synthesis reports best-case benefit--cost ratios of order 300 for \JCA and order 100 for centimeter-class debris removal \cite{Colvin2024IAC}. This is an important portfolio result: permanent removal eliminates all future risk from a target object, but it requires rendezvous, capture, and disposal of a non-cooperative body. A conjunction-specific intervention may capture much of the benefit at materially lower cost if tracking is good enough.

Technically, first-of-a-kind \ADR is now credible. ESA's current ClearSpace-1 mission page describes a 2029 planned demonstration to remove the 95 kg PROBA-1 using a chaser with four robotic arms \cite{ESAClearSpaceCurrent}. JAXA's CRD2 program has already demonstrated close inspection of a non-cooperative H-IIA upper stage with the ADRAS-J spacecraft, including fixed-point observations at about 50 m, and has progressed toward actual removal \cite{JAXACRD2,JAXAADRASJ2024}. Astroscale reports that the CRD2 Phase II contract is worth about 13.2 billion yen including tax \cite{Astroscale2024CRD2}. The engineering challenge is not only capture mechanics. Targets are non-cooperative, may tumble, often lack reliable shape or reflectance models, and were not designed with grapple interfaces. Design-for-removal therefore remains a powerful upstream policy lever \cite{ESAZeroDebris}.

\subsection{Budget-constrained target selection}
\label{sec:budget_selection}

Under a finite remediation budget $B$, target selection becomes a constrained allocation problem:
\begin{equation}
\max_{z_{a,k}\in\{0,1\}}
\sum_{a,k} z_{a,k}\,\eta_{a,k}\Pi_k
\qquad
\text{s.t.}
\qquad
\sum_{a,k} z_{a,k} C_{a,k}\le B,
\label{eq:budget_selection}
\end{equation}
where $z_{a,k}$ indicates whether action $a$ is applied to target $k$, $\eta_{a,k}$ is the fraction of hazard stock removed by action $a$, and $C_{a,k}$ is lifecycle cost. Eq.~(\ref{eq:budget_selection}) makes clear that remediation should be treated as a portfolio selection problem over avoided hazard per dollar, not as a mass-removal or mission-count contest. In practice, this is why a small number of high-$\Pi_k$ rocket bodies or inactive payloads can dominate an entire remediation campaign.

A useful first-pass ranking statistic is
\[
R_{a,k}\equiv \frac{\eta_{a,k}\Pi_k}{C_{a,k}},
\]
which measures avoided hazard stock per dollar. In the fractional-relaxation or weak-interaction limit, larger $R_{a,k}$ should be purchased first; in the exact 0--1 case the optimum remains knapsack-like, but $R_{a,k}$ is still the right engineering screening quantity. This is why remediation should be discussed as a hazard-retirement portfolio rather than as a competition over kilograms removed or missions flown.

\subsection{Hazard-reduction frontier under a finite budget}
\label{sec:frontier}

\begin{figure}[t]
\includegraphics[width=0.58\linewidth]{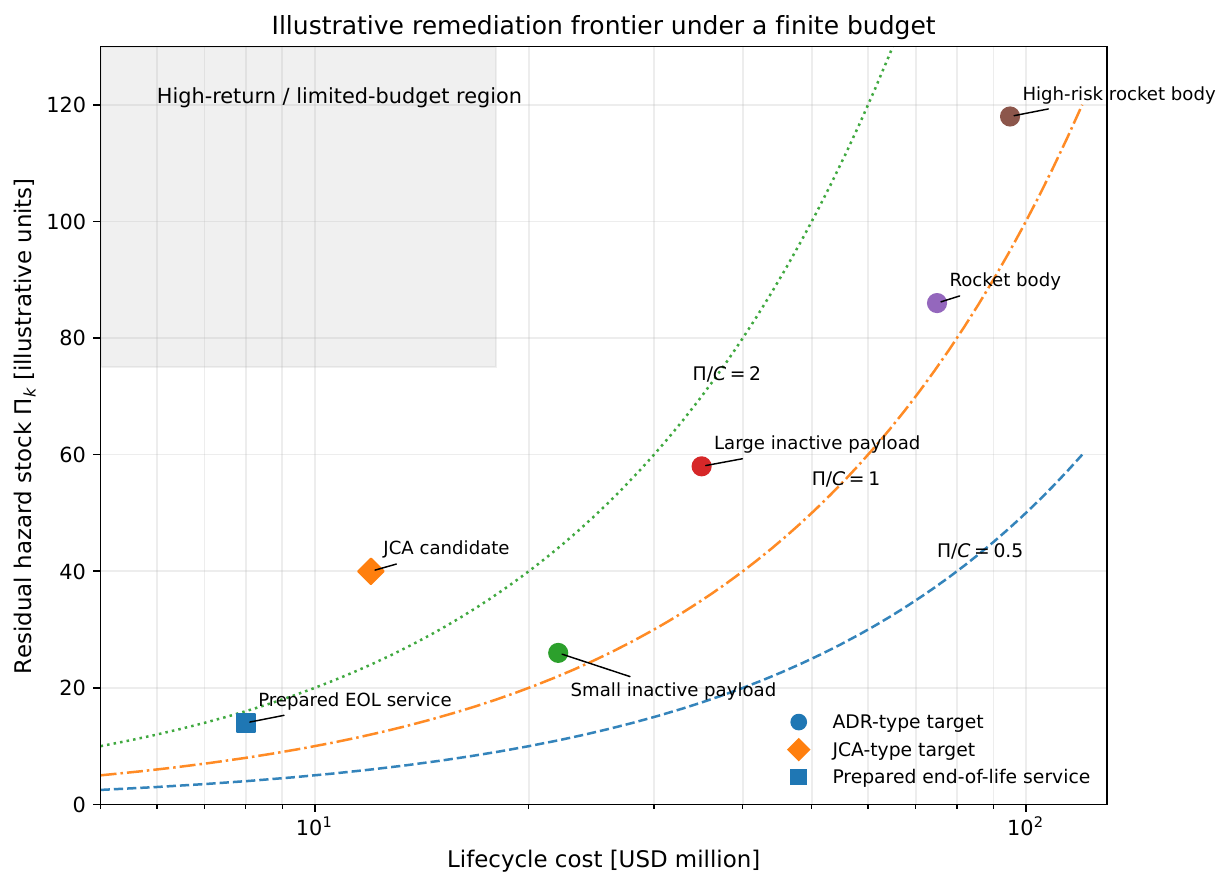}
\vskip -5pt
\caption{Illustrative remediation frontier under a finite budget. Candidate targets are plotted in lifecycle cost versus hazard-stock space, with diagonal contours indicating constant avoided hazard per dollar. The figure is schematic unless calibrated target-level data are available; its role is to visualize the allocation logic of Eq.~(\ref{eq:budget_selection}).}
\label{fig:frontier}
\end{figure}

Eq.~(\ref{eq:budget_selection}) implies that remediation should be organized around avoided hazard per dollar rather than around objects removed per mission. Figure~\ref{fig:frontier} visualizes this by placing candidate targets in a cost--hazard plane and overlaying contours of constant hazard reduction per unit cost. The important qualitative result is that a small set of high-$\Pi_k$ objects can dominate an entire remediation campaign, especially when those objects combine large mass, long residual residence time, and high breakup consequence.

This figure is also where the paper’s intervention logic becomes operationally actionable. If a public buyer or consortium is deciding between \ADR and \JCA, the correct question is not which technology is more impressive, but which candidate action lies furthest toward the upper-left of the cost--hazard plane once legal consent and verification risk are included.

\subsection{Centimeter-class remediation and the limits of direct control}

The 1--10 cm regime is the hardest technical problem. These objects are numerous enough to matter, energetic enough to be lethal, but difficult to track comprehensively with the fidelity required for routine collision avoidance. NASA's phase-2 analysis estimates between roughly 145,000 and 425,000 pieces of 1--10 cm debris in \LEO, depending on the model adopted \cite{NASAOTPS2024}. In optimistic cases centimeter-class remediation can have positive net benefits, but the economics are highly sensitive to model assumptions and infrastructure cost \cite{NASAOTPS2024,Colvin2024IAC}.

A particularly important operational caveat is that tracking centimeter-class debris only helps if state uncertainties are much smaller than those associated with today's trackable objects. NASA warns that if conjunction uncertainties for centimeter-class debris are not substantially better, operators will be inundated with warnings and maneuvers, driving net benefits negative \cite{NASAOTPS2024}. Consequently centimeter-class tracking or remediation should not be treated as the first-line response for today's environment. It is a second-order capability that becomes valuable only after rapid disposal, better high-risk tracking, and selective large-object intervention are operating at scale.

\subsection{Shielding and resilient design as local controls}

Shielding is necessary, but it is a local measure rather than an environmental one. NASA's HyperVelocity Impact Technology group describes the logic of Whipple and stuffed-Whipple concepts: a sacrificial bumper fragments and disperses the projectile, while intermediate layers further shock and pulverize the debris cloud before it reaches the rear wall \cite{NASAHVIT}. NASA's debris FAQ notes that the ISS is the most heavily shielded spacecraft ever flown and that critical components are generally designed to withstand impacts by debris as large as 1 cm in diameter \cite{NASAODPOFAQ}. Robotic spacecraft rarely have ISS-class mass margins; the mass cost of protection rises quickly with required protection level. In practice shielding should be used to protect only vulnerability-sensitive surfaces or subsystems while environmental controls suppress the source population.

\subsection{Operator-facing metrics and disclosure standards}
\label{sec:kpis}

For industry use, the debris problem should be monitored through a compact disclosure set rather than through narrative reporting alone. A minimum useful set is
\begin{align}
\CAR(P_c^\star) &= \frac{M_{\rm CA}}{N_{\rm sat}\,\Delta t}, &
\OYF &= \sum_{k \in \mathcal{F}} \tau_k^{\rm passive}, &
\DMR_{5} &= \frac{N_{\rm deorb}(<5\,{\rm yr})}{N_{\rm retired}},
\label{eq:KPIs1} \\
\ERI &= \mathrm{median}\!\left(\sqrt{\det C_b}\right)_{\rm high\ risk}, &
\AOU &= \frac{T_{\rm outage}}{\Delta t},
\label{eq:KPIs2}
\end{align}
where $M_{\rm CA}$ is the number of executed collision-avoidance maneuvers in reporting interval $\Delta t$, $N_{\rm sat}$ is the mean active spacecraft count over the same interval, $\mathcal{F}$ is the set of failed or passively drifting spacecraft, $\tau_k^{\rm passive}$ is the passive post-failure residence time of object $k$, $N_{\rm deorb}(<5\,{\rm yr})$ is the number of retired spacecraft successfully deorbited within 5 years, $N_{\rm retired}$ is the total number of retired spacecraft in the reporting interval, and $T_{\rm outage}$ is the time during which the collision-avoidance or conjunction-support chain is unavailable. Here $\ERI$ is an encounter-risk inflation metric computed over the operator-defined high-risk conjunction subset.

Such metrics are more useful than simple launch counts or raw maneuver counts because they separate environmental burden, operator policy, and system reliability. They also provide a practical bridge between operations and finance. In particular, $\CAR(P_c^\star)$, $\DMR_5$, $\OYF$, and $\ERI$ can serve as auditable inputs to disposal bonds, performance-based licensing, insurance differentiation, remediation target ranking, and service-procurement verification. A debris-control market cannot support durable private investment if the underlying operational state variables are not reported in a form that is comparable across operators and over time.

For comparability across operators, the KPI set defined by Eqs.~(\ref{eq:KPIs1})--(\ref{eq:KPIs2}) is only meaningful if accompanied by a minimum metadata set: (i) the maneuver trigger threshold $P_c^\star$; (ii) the reporting interval $\Delta t$; (iii) the definition of active spacecraft used for $N_{\rm sat}$; (iv) the fraction of maneuvers executed autonomously versus manually approved; (v) the screening horizon and covariance source used in conjunction assessment; and (vi) the definition of retired, failed, and passively drifting states used in computing $\DMR_5$ and $\OYF$. Without this metadata, public operator reporting remains informative but not analytically comparable.

\section{NASA, public infrastructure, and industry implementation}
\label{sec:implementation}

\subsection{NASA as a technical actor, not merely a citation source}

NASA should appear in this problem not only as a data provider but as a central technical actor. NASA's Orbital Debris Program Office explicitly organizes its work across modeling, measurements, protection, mitigation, remediation, and reentry \cite{NASAODPOMain}. On the modeling side, ORDEM~3.2 is NASA's engineering debris-flux model for spacecraft design and impact-risk assessment, and LEGEND is the agency's primary three-dimensional evolutionary model for projecting the long-term debris environment and evaluating mitigation or remediation strategies \cite{LEGENDNASA}. DAS~3.2 translates this knowledge into mission-level compliance and design analysis \cite{DAS32}. NASA updated NPR~8715.6E in 2024 as the current agency procedural requirement for orbital debris mitigation \cite{NPR87156E}.

NASA has also acted directly in the regulatory and policy layer. In the 2022 Starlink Gen2 proceeding, NASA warned that deployment should be conducted prudently to support spaceflight safety, science missions, and the long-term sustainability of the orbital environment \cite{NASAFCCStarlink2022}. FCC documentation in 2024 and 2026 later referred to continuing NASA--SpaceX coordination on autonomous collision avoidance and coexistence with ISS operations \cite{FCCGen2LowerAlt2024,FCC2026SpaceXThreshold}. Finally, NASA's Office of Technology, Policy, and Strategy has extended the agency's role from technical analysis into cost-benefit studies of mitigation, tracking, remediation, and portfolio design \cite{NASAOTPS2023,NASAOTPS2024,Rao2025Portfolio}. NASA is therefore not merely observing the debris problem; it is shaping the measurement, modeling, procedural, and economic layers through which the orbital environment is controlled.

The NASA stack maps naturally onto the three-control framework used in this paper. ORDEM constrains local debris-flux exposure and therefore informs shielding, vulnerability, and design margins. LEGEND resolves the long-horizon sensitivity of the environment to assumptions on \PMD, \ADR, and breakup source terms and therefore informs the residual hazard stock $\Pi_k$. DAS and NPR~8715.6E translate mitigation requirements into mission design and compliance, acting primarily on disposal reliability $p_{\rm disp}$. The conjunction-assessment handbook and related operational interfaces act on encounter-state uncertainty through $C_b$. NASA's role is therefore coherent across the full control chain: measurement, model development, design guidance, procedural requirements, and economic evaluation. 

\subsection{Public \SSA/\STM infrastructure and TraCSS}

The U.S. civil \SSA/\STM architecture is also evolving. NOAA's Office of Space Commerce is developing the Traffic Coordination System for Space (TraCSS) to provide baseline civil \SSA data and services to private and civil operators. As of February 2026, TraCSS had 17 pilot users, and NOAA had opened a waitlist as the system moved toward fuller production \cite{TraCSS2026,TraCSSWaitlist2026}. This is an important institutional development because it moves civil conjunction support away from ad hoc dependence on defense interfaces and toward a public utility model with commercial participation.

The correct role of such infrastructure is not full state ownership of all sensing or screening functions. It is provision of a neutral baseline layer---catalog access, conjunction screening, user authentication, and common data standards---onto which commercial data providers and high-precision tasking services can be layered. In economic terms, baseline civil \SSA/\STM behaves like a public or club good with strong network externalities, while premium covariance improvement and special tasking can operate as commercial overlays.

The resulting architecture is layered rather than exclusive: government provides the neutral baseline utility, while commercial actors compete on covariance improvement, sensor tasking, analytics, autonomy support, and operator-specific integration.

\subsection{Industry transparency, operator reporting, and servicing markets}

Industrial execution matters because many of the relevant controls are operational rather than conceptual. Public Starlink maneuver reporting has already shown the value of operator-level transparency, even though the current reporting regime remains incomplete and the metrics themselves require normalization \cite{SpaceXStarlink2022,SpaceXSemiannual2025Jul,StarlinkAerospaceAmerica2025}. The same lesson applies to disposal reliability and system outages: without structured disclosure, external oversight cannot distinguish a challenging environment from poor operational practice.

Prepared servicing is presently the least institutionally constrained operator-procured debris-control service class. Astroscale announced in June 2025 that ELSA-M completed critical design review, targets a 2026 launch, and is majority privately funded, with co-funding from the UK Space Agency via ESA and from Eutelsat \cite{AstroscaleELSAMCDR2025}. The relevant point is not ``commercial'' appeal in the abstract, but counterpart structure: prepared satellites provide identifiable customers, pre-agreed interfaces, and auditable performance. By contrast, remediation of unprepared legacy derelicts is not well suited to decentralized bilateral procurement because both benefit incidence and legal authority are diffuse.

\section{Economics, financing, and allocation of responsibility}
\label{sec:econ}

\subsection{Externality and dynamic portfolio optimization}

The debris problem is a textbook externality with delayed and probabilistic payoffs. The operator who pays for improved disposal hardware or higher-fidelity tracking does not capture all of the benefit; much of the gain accrues to other operators and often arrives years later. NASA's 2023 debris-remediation analysis states directly that the substantial up-front expenditures and delayed benefits do not appear sufficient, on their own, to incentivize immediate private action \cite{NASAOTPS2023}. The same study argues that reductions in risk to any single entity are likely too small to motivate that entity to purchase remediation services at the socially optimal level.

This is not an argument against markets; it is an argument that the market must be structured. Let $\mathbf u(t)$ collect mitigation, tracking, and remediation controls, let $T$ denote the planning horizon, and let $r$ denote the social discount rate. The appropriate welfare problem is
\begin{equation}
\min_{\mathbf u(t)}
\int_0^T e^{-rt}\left[\Dmg(\mathbf N(t),\mathbf u(t)) + C(\mathbf u(t))\right]dt
\qquad
\text{s.t.}\qquad
\dot{\mathbf N}=\mathbf f(\mathbf N,\mathbf u,t),
\label{eq:portfolio_problem}
\end{equation}
where $\Dmg$ is expected debris damage and $C$ is intervention cost. Eq.~(\ref{eq:portfolio_problem}) implies that rapid \PMD, targeted uncertainty reduction, and selective \ADR are not independent substitutes; they are interacting controls whose marginal values depend on the orbital state and on which other controls are already active. NASA's 2025 follow-on work makes the same point: efficient debris-risk policy requires portfolio thinking rather than isolated benefit--cost ratios \cite{Rao2025Portfolio}.

\subsection{Allocation of responsibility and implementation mechanisms}

A rigorous allocation of responsibility follows from benefit incidence, excludability, and legal control over the relevant object. Let $x$ denote a debris-control action, $B_i(x)$ the benefit to operator $i$, and $C(x)$ the cost of the action. The core institutional problem is that the privately captured marginal benefit is often smaller than the socially relevant marginal benefit:
\begin{equation}
\Delta_x \equiv \frac{d}{dx}\sum_{i=1}^{N} B_i(x) - \frac{d}{dx}B_f(x) > 0,
\label{eq:externality_gap}
\end{equation}
where $B_f$ is the benefit captured by the party expected to pay. When $\Delta_x$ is large, the action is underprovided absent regulation, procurement, subsidies, or transfer mechanisms. The central economic problem is to reduce the externality gap between the benefit captured by the funding party and the larger benefit distributed across all operators.

The legal structure reinforces the need for public involvement. Under Article VIII of the Outer Space Treaty and the Registration Convention, the state of registry retains jurisdiction and control over a space object \cite{OuterSpaceTreaty1967,RegistrationConvention1975}. Legacy \ADR is therefore not a conventional salvage market. Intervention on an unprepared object usually requires state consent, contractual authorization, or both. The efficient institutional division is asymmetric rather than uniform. Operators should internalize the cost of not creating future debris from their own missions. Governments should fund or procure common-good infrastructure and at least the first wave of legacy-object remediation. Between these endpoints lies an intermediate procurement regime---prepared end-of-life servicing, consortium high-fidelity tracking, and public purchase of operator-specific conjunction services---for which counterparties are identifiable and performance variables are contractible \cite{TraCSS2026,AstroscaleELSAMCDR2025}.

\begin{table*}[t]
\caption{Control stack for orbital debris: dominant physical effect, preferred payer, and leading implementation instrument.}
\label{tab:controlstack}
\begin{tabular}{l l l l l}
\toprule
{Layer} & {Main physical effect} & {Preferred payer} & {Leading instrument} & {When it dominates} \\
\midrule
New-mission mit- &
Reduces future source  &
Private operator  &
Performance standard  &
All new missions in crowded  \\
igation / \PMD &
 term and post-failure  &
 under license  & plus deposit-refund or  &
\LEO shells \\
 & residence time & conditions &
performance bond & \\[2pt]

Targeted  &
Reduces encounter un- &
Public utility with  &
Civil baseline service plus  &
Small subset of conjunctions \\

 \SSA/\STM &
certainty, false alarms,  &
 commercial overlays &
commercial tasking for &
 that dominates operational  \\
& and maneuver burden & & high-risk conjunctions &
risk \\[2pt]

Legacy &
Reduces residual haz- &
Public procure- &
Outcome-based remedia- &
Persistent high-risk rocket  \\

 \JCA/\ADR &
ard stock of inactive  &
ment or pooled  &
tion contracts tied to ver-&
 bodies and inactive payloads \\

 & high-mass objects &
 consortium finance &
ified risk reduction &  \\[2pt]

Shielding / re- &
Reduces local vulnera- &
Mission operator &
Mission design and sub- &
Residual mm--cm environ-  \\

silient design &
bility, not environ& &
system hardening &
ment that cannot be econom- \\

 & mental hazard & &
& ically controlled upstream \\
\bottomrule
\end{tabular}
\end{table*}

Table~\ref{tab:controlstack} implies that orbital-debris control separates into distinct procurement regimes rather than a single uniform sector. New-mission mitigation and prepared end-of-life intervention have identifiable counterparties and largely appropriable benefits and can therefore be procured directly by operators. Baseline civil \SSA/\STM retains public-infrastructure characteristics, with operator-specific analytics and tasking layered on top. Remediation of unprepared legacy derelicts belongs primarily to a publicly procured regime because the benefit is diffuse and the relevant authority is state-mediated.

What already exists is uneven. Command-and-control obligations already force private spending on debris mitigation for new missions. Public procurement of commercial remediation capability already exists through ESA and JAXA. Public civil \SSA/\STM utilities are emerging through TraCSS. What does \emph{not} yet exist in mature form is a complete mechanism for pricing long-lived post-mission risk or paying private actors for verified reduction of environmental hazard. The most important missing instruments are therefore those that translate environmental harm into private cost or translate verified environmental benefit into revenue.

A conceptually clean ex ante charge is a risk-weighted orbit-use fee tied to an object's expected marginal contribution to the hazard functional:
\begin{equation}
F_k^{\rm orb} = \alpha\,\Pi_k^{\rm proxy},
\qquad
\Pi_k^{\rm proxy} \propto m_k\,\tau_k\,\phi_k\,\chi_k\,\big(1-p_{{\rm disp},k}\big),
\label{eq:risk_fee}
\end{equation}
where $m_k$ is object mass, $\tau_k$ is expected post-mission residence time in protected shells, $\phi_k$ is a shell-dependent collision-flux proxy, $\chi_k$ is a breakup-severity factor, and $p_{{\rm disp},k}$ is verified disposal success probability. In practice this would be implemented through observable proxies such as altitude, mass, cross-section, maneuverability, and demonstrated disposal reliability rather than through a full environment simulation at license issue. OECD's 2024 review is explicit that incentive-based measures such as launch or orbital taxes and performance bonds can produce positive long-term environmental and efficiency gains, although coordination and competitiveness remain open design problems \cite{OECD2024Sustainability}.

Eq.~(\ref{eq:risk_fee}) should be interpreted as an administrative approximation to the true shadow price of environmental harm. It is not intended to reproduce a full evolutionary debris simulation at licence issue. Its purpose is to align ex ante private cost with the leading physical drivers of long-run hazard: intact mass, residence time in protected shells, local collision flux, breakup consequence, and disposal failure probability. In practice, the credibility of such a fee depends less on model sophistication than on transparent calibration and predictable refund or bond-release rules.

For market design, Eq.~(\ref{eq:risk_fee}) implies that pricing must be based on variables that are both physically relevant and administratively observable. In the near term, mass, orbital altitude, expected residence time, maneuverability, and verified disposal reliability are therefore more useful pricing inputs than model sophistication alone.

\paragraph*{Illustrative pricing arithmetic.}
Eq.~(\ref{eq:risk_fee}) is most useful if it can distinguish low-persistence, low-liability missions from long-lived, high-consequence objects. Suppose a regulator adopts a normalization in which
\[
F_k^{\rm orb} = \alpha \Pi_k^{\rm proxy},
\qquad
\Pi_k^{\rm proxy} \propto m_k \tau_k \phi_k \chi_k \big(1-p_{{\rm disp},k}\big),
\]
with $\alpha$ chosen so that a rapidly disposable low-altitude mission faces a modest fee. Table~\ref{tab:riskfee_example} then illustrates the qualitative ordering implied by Eq.~(\ref{eq:risk_fee}): persistence, intact mass, and disposal failure probability can dominate the fee signal even before a full calibrated environment shadow price is available.

Consider three representative cases: (i) a 300 kg spacecraft at 550 km with high disposal reliability, (ii) a 2,000 kg spacecraft at 850 km with only moderate disposal reliability, and (iii) an upper-stage-class object of similar altitude and substantially higher breakup consequence. Even without assigning absolute fee values, Eq.~(\ref{eq:risk_fee}) implies a steep ordering in expected ex ante charge, driven primarily by persistence, intact mass, and disposal failure probability. The practical consequence is that pricing can be made directionally correct using variables that are already observable at licensing, even before a fully calibrated environment shadow price is available. Table~\ref{tab:riskfee_example} illustrates these cases.

\begin{table}[t]
\caption{Illustrative ordering implied by Eq.~(\ref{eq:risk_fee}). Values are normalized and schematic.}
\label{tab:riskfee_example}
\begin{tabular}{lccc}
\toprule
Case & $m_k$ & $\tau_k$ & Relative $F_k^{\rm orb}$ \\
\midrule
550 km spacecraft, high $p_{\rm disp}$ & low & low & 1 \\
850 km spacecraft, moderate $p_{\rm disp}$ & medium & high & 8 \\
850 km upper-stage-class object & high & high & 20 \\
\bottomrule
\end{tabular}
\end{table}

\subsection{Expected-value conditions for privately financed service provision}
\label{sec:commercial_viability}

A debris-control service can support privately financed provision only when the deliverable is observable, verifiable, and legally executable on the relevant target set. For a service project $s$, a necessary expected-value condition is
\begin{equation}
\mathbb{E}[\mathrm{NPV}_s]
=
p_{\rm award}\,p_{\rm consent}\,p_{\rm verify}\,P_s
-
C^{\rm dev}_s
-
C^{\rm ops}_s
-
C^{\rm liab}_s
-
C^{\rm cap}_s
> 0,
\label{eq:npv_service}
\end{equation}
where $P_s$ is contracted payment, $p_{\rm award}$ is award probability, $p_{\rm consent}$ is the probability of legal authority to act on the target, $p_{\rm verify}$ is the probability that contract performance is verified, and the $C_s$ terms denote development, operations, liability, and capital costs. In what follows, private provision is feasible when Eq.~(\ref{eq:npv_service}) is positive under the stated award, consent, verification, payment, and cost assumptions.

Eq.~(\ref{eq:npv_service}) implies four implementation conditions. First, the procured output must be measurable and verifiable; examples are verified disposal, verified reduction in passive object-years, or verified reduction in residual hazard stock. Second, large spacecraft and upper stages must be serviceable by design, so that intervention is technically and contractually tractable. Third, liability allocation and state-consent procedures must be standardized, especially for unprepared legacy targets. Fourth, early demand for legacy-stock remediation must be supplied through public or multilateral procurement, because the benefit is diffuse and cannot usually be appropriated by a single operator.

Eq.~(\ref{eq:npv_service}) also shows that technical feasibility is not sufficient for private provision. The decisive uncertainties may be the realization probabilities $p_{\rm award}$, $p_{\rm consent}$, and $p_{\rm verify}$ rather than rendezvous or deorbit capability alone. The implementation problem is therefore jointly technical, legal, and contractual.

Eq.~(\ref{eq:npv_service}) becomes more decision-relevant when inverted:
\begin{equation}
P_s^{\rm min}
=
\frac{C^{\rm dev}_s+C^{\rm ops}_s+C^{\rm liab}_s+C^{\rm cap}_s}
{p_{\rm award}\,p_{\rm consent}\,p_{\rm verify}},
\label{eq:min_payment}
\end{equation}
which is the minimum contract value required for nonnegative expected NPV. Let
\[
p_{\rm real}\equiv p_{\rm award}\,p_{\rm consent}\,p_{\rm verify}
\]
denote the joint realization probability. Then Eq.~(\ref{eq:min_payment}) can be read as
\[
P_s^{\rm min}=\frac{C_s^{\rm tot}}{p_{\rm real}},
\qquad
C_s^{\rm tot}\equiv
C^{\rm dev}_s+C^{\rm ops}_s+C^{\rm liab}_s+C^{\rm cap}_s .
\]
Thus $P_s^{\rm min}$ is the minimum contracted payment compatible with nonnegative expected net present value. In the two illustrative cases below, $p_{\rm real}=0.855$ for prepared end-of-life service and $p_{\rm real}=0.448$ for unprepared legacy remediation, implying $P_s^{\rm min}\simeq 26.9$ and $91.5$, respectively. The second case therefore requires about \(0.855/0.448 \simeq 1.9\) times larger contracted value solely because the joint realization probability is lower.

A useful way to translate Eq.~(\ref{eq:npv_service}) into contract design is to express payment as a mixture of fixed and outcome-based terms:
\begin{equation}
P_s
=
P^{\rm fix}_s
+
\beta_{\Pi}\,\Delta \Pi^{\rm ver}_s
+
\beta_{\rm disp}\,N^{\rm ver}_{{\rm disp},s}
+
\beta_{\rm JCA}\,N^{\rm ver}_{{\rm JCA},s}
-
L_s,
\label{eq:payment_formula}
\end{equation}
where $P^{\rm fix}_s$ is the fixed mobilization or readiness payment, $\Delta \Pi^{\rm ver}_s$ is verified reduction in hazard stock, $N^{\rm ver}_{{\rm disp},s}$ is the number of verified disposals completed, $N^{\rm ver}_{{\rm JCA},s}$ is the number of verified avoided conjunction events, and $L_s$ collects penalties, abatements, or claw-backs associated with non-performance or verification failure.

Eq.~(\ref{eq:payment_formula}) is useful because it makes explicit that different market segments require different dominant payment terms. Compliance-led \PMD markets will tend to be dominated by $\beta_{\rm disp}$ and bond-release logic. Premium \SSA/\STM services will tend to be dominated by fixed or subscription revenue plus event-based tasking. Legacy remediation will tend to require a larger fixed readiness payment together with an outcome-based hazard-reduction term $\beta_{\Pi}\,\Delta \Pi^{\rm ver}_s$. 
Eq.~(\ref{eq:payment_formula}) also implies that procurement should not be written around ``objects removed'' as the principal deliverable. The contractible product should instead be verified disposal, verified avoided conjunctions, or verified reduction in hazard stock, as those are the quantities that map directly back to the physical control variables of the orbital environment.

\subsection{Illustrative service-case examples}
\label{sec:service_case}

The commercial-viability condition in Eq.~(\ref{eq:npv_service}) is easier to interpret when translated into stylized service arithmetic for two representative market segments.

\paragraph*{Case A: prepared end-of-life servicing.}
Suppose a provider has a fleet contract with
\[
P_s = P_s^{\rm fix} + \beta_{\rm disp}N^{\rm ver}_{{\rm disp},s}
= 8 + 24 = 32,
\]
with $p_{\rm award}=0.90$, $p_{\rm consent}=1.0$, and $p_{\rm verify}=0.95$. If
\[
C_s^{\rm dev}=6,\qquad
C_s^{\rm ops}=14,\qquad
C_s^{\rm liab}=2,\qquad
C_s^{\rm cap}=1,
\]
then
\begin{equation}
\mathbb{E}[\mathrm{NPV}_s]
=
0.90\times 1.0\times 0.95 \times 32 - (6+14+2+1)
\simeq 4.4.
\end{equation}

This positive value indicates that prepared end-of-life servicing is the least institutionally constrained early operator-procured service class under the stated assumptions: the counterparty is identifiable, consent is intrinsic to the contract, and the realization probability remains high.

\paragraph*{Case B: unprepared legacy remediation.}
Now consider a legacy remediation service with
\[
P_s = P_s^{\rm fix} + \beta_{\Pi}\Delta \Pi_s^{\rm ver}
= 25 + 40 = 65,
\]
but with $p_{\rm award}=0.70$, $p_{\rm consent}=0.80$, and $p_{\rm verify}=0.80$, reflecting consent complexity and outcome-verification risk. If
\[
C_s^{\rm dev}=10,\qquad
C_s^{\rm ops}=22,\qquad
C_s^{\rm liab}=5,\qquad
C_s^{\rm cap}=4,
\]
then
\begin{equation}
\mathbb{E}[\mathrm{NPV}_s]
=
0.70\times 0.80\times 0.80 \times 65 - (10+22+5+4)
\simeq -11.9.
\end{equation}
The stylized result is negative despite the larger nominal contract value because the realization probability is lower and the cost base is larger. Under these assumptions, legacy-stock remediation is not compatible with decentralized private procurement; it instead requires public or consortium procurement, standardized consent procedures, and auditable verification of hazard reduction.

\subsection{Service segmentation implied by the control architecture}
\label{sec:market_architecture}

Table~\ref{tab:controlstack} and Eq.~(\ref{eq:npv_service}) imply service segmentation rather than a single homogeneous debris-services sector. The relevant discriminants are customer specificity, excludability, benefit diffuseness, and consent complexity. Services tied to new-mission disposal and prepared end-of-life intervention have identifiable counterparties and largely appropriable benefits and can therefore be procured directly by operators. Civil baseline \SSA/\STM retains public-infrastructure characteristics. Remediation of unprepared legacy objects differs because benefits are diffuse and legal authority is state-mediated.

\begin{figure}[t]
\includegraphics[width=0.55\linewidth]{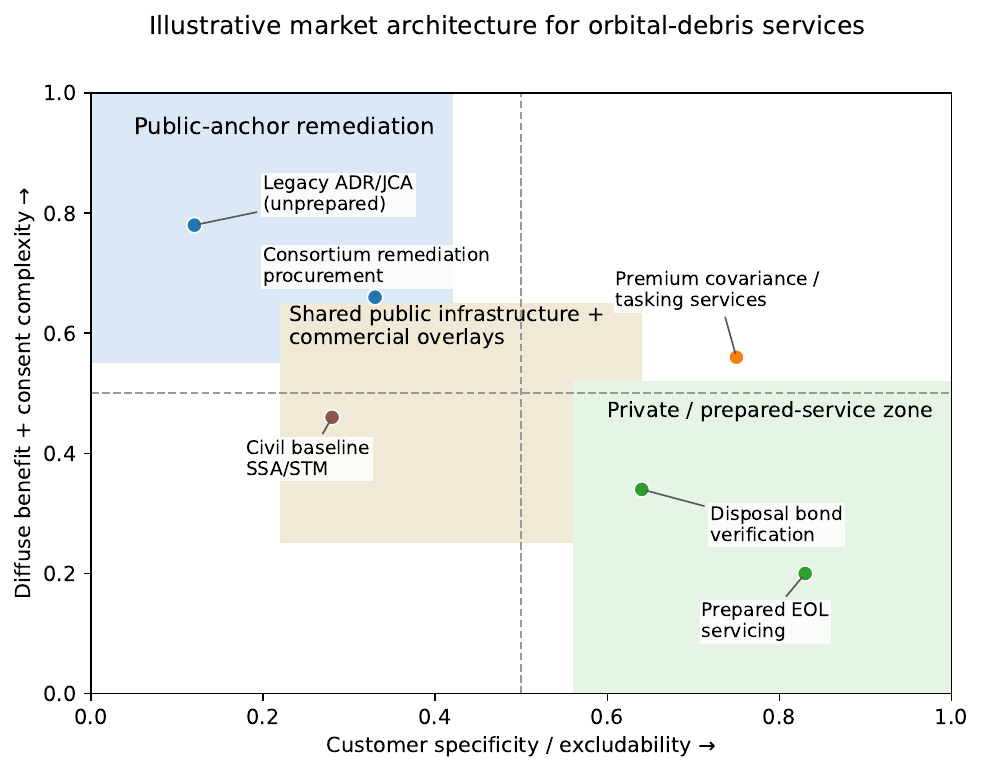}
\vskip-5pt 
\caption{Illustrative procurement map for orbital-debris control. The horizontal axis denotes customer specificity / excludability; the vertical axis denotes benefit diffuseness and consent complexity. Prepared end-of-life servicing lies in the operator-procured regime, whereas remediation of unprepared legacy objects lies in the publicly procured regime.}
\label{fig:market_map}
\end{figure}

The practical consequence is not a generic claim about ``markets,'' but a classification of procurement regimes. Prepared end-of-life servicing lies in the operator-procured regime because performance is contractible and consent is intrinsic to the engagement. Legacy-stock remediation lies in the publicly procured regime because the socially relevant benefit exceeds the benefit appropriable by any single operator. Figure~\ref{fig:market_map} summarizes this separation.

\subsection{Sequential implementation priorities}
\label{sec:market_scenarios}

The control architecture developed here implies a sequence of procurement regimes rather than simultaneous emergence of a single debris-services sector.

\paragraph*{New-mission mitigation.}
The counterparty is the operator, the deliverable is verified disposal capability or verified disposal outcome, and the dominant control variable is $p_{\rm disp}$.

\paragraph*{Prepared servicing and covariance-reduction services.}
The counterparty remains identifiable, the deliverable is operational---prepared end-of-life servicing, covariance reduction, precision tasking, or conjunction-decision support---and the dominant control variables are $p_{\rm disp}$ and $C_b$.

\paragraph*{Legacy-stock remediation.}
For unprepared legacy objects, the appropriate deliverable is verified reduction in residual hazard stock $\Pi_k$. Because benefits are diffuse and consent is state-mediated, this regime is naturally organized through government or multilateral procurement with commercial execution.

This sequencing is not rhetorical. Each layer reduces uncertainty in the next: audited disposal improves reporting and liability allocation; standardized interfaces and covariance products reduce execution uncertainty for operator-procured services; and only then does large-scale legacy remediation become suitable for sustained procurement rather than isolated demonstrations.

\begin{table*}[t]
\caption{Contract structure and implementation constraints for orbital-debris service classes.}
\label{tab:business_cases}
\begin{tabular}{l l l l l l}
\toprule
{Market segment} & {Natural buyer} & {Unit of sale} & {Revenue model} & {Verification basis} & {Primary blockers} \\
\midrule
Compliance-led  &
Operator / con- &
Verified disposal ca- &
Hardware sale, com-&
$\DMR_5$, verified  &
Weak enforcement,  \\

mitigation/\PMD &
stellation owner &
pability or verified  &
pliance service fee,  &
disposal, opera- &
inconsistent stan-\\
& & disposal outcome &  refundable bond re- &
tor reporting & dards, end-of-life fail- \\
& & & lease, insurance pre- &  & ure risk \\
& & & mium reduction & & \\

Prepared end-of- &
Operator / fleet  &
Serviced spacecraft &
Per-spacecraft con- &
Verified docking,  &
Lack of standard  \\

life servicing & owner & or verified end-of- &
tract, fleet subscrip- &
towing, disposal  &
 interfaces, limited \\
 &  & life removal & tion, milestone +  & completion &
prepared-target base \\
&  & & success fee &  &  \\

Premium  &
Operator, in- &
Covariance improve&
Subscription, event-&
$\ERI$, documen-&
Fragmented data  \\

 \SSA/\STM  &
surer, civil  &
ment, tasking event,    &
based tasking, pre- &
ted covariance ,  &
access, inconsistent  \\

services & agency & screening subscrip-  & mium analytics &
improvement  &
standards, dependen- \\

 &  &  tion  &  &
 task completion  &
ce on public baseline \\

Legacy  &
Government,  &
Verified: reduction  &
Public procurement, &
$\Delta \Pi_k^{\rm ver}$, verified  &
Consent, liability,\\

 \JCA/\ADR for un- &
agency, multi- &
in $\Pi_k$, avoided con- &
milestone + out- &
removal, verified &
target uncertainty, \\

prepared targets &
lateral  consor- & junction,  removal of & come payment &
avoidance event & lumpy public demand \\

 & tium & named target &  & & \\

Data / verifica- &
Regulator, in- &
Audit of KPIs,  &
Audit contract, cer-&
Independent ver- &
Lack of trusted stan- \\

tion / audit layer &
surer, procure-&
bond release, perfor- &
tification fee, com-&
ification of KPIs  &
dards, no accepted \\

& ment authority  & mance certification&
pliance support  &
 and outcomes &
audit regime\\

\bottomrule
\end{tabular}
\end{table*}

\subsection{What current cost-benefit work implies}

Table~\ref{tab:econ} is best read as a marginal-value ordering, not as a menu of disconnected ratios. The consistent pattern across the public studies cited here is: first reduce $(1-p_{\rm disp})\tau_k$ for new missions, then reduce $\sqrt{\det C_b}$ for the conjunctions that dominate operator workload, then selectively retire the highest-$\Pi_k$ legacy stock, and only after that scale centimeter-class custody. The common reason is implementation leverage: the first three actions attack the dominant terms in the reduced-order control architecture with much lower sensing, rendezvous, and verification complexity than broad cm-class control \cite{NASAOTPS2024,Colvin2024IAC,NASAOTPS2023}.

\begin{table*}[t]
\caption{Selected public technoeconomic results relevant to debris policy and technology prioritization. Reported ranges are scenario-conditioned decision support values rather than project-level financial returns.}
\label{tab:econ}
\begin{tabular}{l l l }
\toprule
{Action} & {Published benefit or ratio} & {Interpretation} \\
\midrule
Change \PMD requirement from  & 20--750 times cost; up to about \$6B  & Strong support for shortening disposal time- \\
25 years to 15 years &  net benefit over 30 years & lines \cite{NASAOTPS2024,Colvin2024IAC} \\

Move from 25-year disposal to  & Nearly \$9B net benefit over 30 years & Shows that faster clearance continues to add  \\
 0-year rule &  & value \cite{NASAOTPS2024,Colvin2024IAC} \\

Deorbit in 5 years or fewer & Benefits exceed costs by several & Rapid clearance is economically dominant \\
&  hundred times & for many new spacecraft \cite{Colvin2024IAC} \\

5-year rule implemented with  & Potentially about $10^3$ times cost in & Very high upside, but wide-area devices in- \\
 with drag devices & optimistic case & crease collision cross-section \cite{Colvin2024IAC}\\

On-demand uncertainty reduc- & Benefits $>100$ times cost; 10x uncer- & Highest-value \SSA/\STM improvement is tar- \\
tion for high-risk conjunctions &tainty reduction worth about \$1.5B & geted high-risk tracking \cite{NASAOTPS2024,Colvin2024IAC} \\
 & over 30 years &  \\

\JCA for large derelicts & Up to $\sim$ 300 times cost in best case & Can be more efficient than full removal in \\
&  & some scenarios \cite{Colvin2024IAC} \\

Centimeter-class debris removal & Up to $\sim$  100 times cost in best case & Promising, but more uncertain than \PMD \\
&  & or targeted tracking \cite{Colvin2024IAC} \\

Remove 50 large debris objects & Roughly 0.1 to 10 times cost after  & Feasible if targets and cost structure are \\

& 30 years & chosen well; not universally dominant \cite{NASAOTPS2024}  \\

\bottomrule
\end{tabular}
\end{table*}

\section{Priority program for 2026--2035}
\label{sec:program}

The engineering objective over the next decade should be to drive the environment away from the unstable regime identified by ESA while reducing operational cost growth in the most congested \LEO shells. A practical program follows directly from the control variables identified above.

The six program elements below map directly onto the three control variables identified in Secs.~\ref{sec:physics} and \ref{sec:interventions}. Items 1 and 6 act primarily on disposal reliability $p_{\rm disp}$ for newly launched spacecraft. Item 2 acts primarily on encounter-state uncertainty through the combined covariance $C_b$ in the high-risk conjunction tail. Items 3 and 4 act primarily on the residual hazard stock $\Pi_k$ of legacy objects, with item 3 targeting large inactive derelicts and item 4 reserving centimeter-class intervention for a later, more mature regime. Item 5 acts locally on spacecraft vulnerability rather than on the environment itself.
\paragraph*{1. Make rapid disposal the default design rule for all new crowded-\LEO missions.}
All new spacecraft in dense \LEO shells should be designed for disposal in less than 5 years after end of mission, and substantially faster in the most populated bands when feasible. Disposal architecture should be reliability-driven: propulsive clearance with margin if available, passive drag augmentation as backup, autonomous failure handling, and verifiable passivation. Large spacecraft and rocket bodies should include serviceable or grappling interfaces to reduce future remediation cost.

\paragraph*{2. Build a high-fidelity targeted tracking layer around existing catalogues.}
The next major operational improvement is not a monolithic global sensor rebuild, but a targeted high-fidelity layer for risky conjunctions. This includes better public/private ephemeris sharing, on-demand tasking of laser ranging and precision optical/radar sensors, better covariance realism, and automated coordination logic. The economic evidence for this layer is unusually strong, and the required technologies already exist in partial operational form.

\paragraph*{3. Publicly procure and co-finance remediation of the highest-risk legacy objects.}
Because the dominant environmental risk is concentrated in inactive objects, and because the benefits of their remediation are diffuse, the first wave of large-object intervention should be organized through public procurement or pooled consortium finance rather than through expectation of a purely private spot market. Governments, agencies, or regulated multilateral consortia should define target lists using transparent risk-ranking metrics such as Eqs.~(\ref{eq:Pi_stock}) and (\ref{eq:risk_fee}), then buy verified risk reduction from competing service providers. Verification should be outcome-based: the relevant product is not kilograms removed, but reduction in expected future fragment generation from the identified target set.

\paragraph*{4. Do not overinvest early in centimeter-class custody.}
Centimeter-class tracking and removal can become valuable, but only after rapid \PMD, high-risk tracking, and legacy derelict interventions are operating at scale. The reason is technical and economic: centimeter-class systems need much better uncertainties and much larger sensing or engagement infrastructure to produce positive net value.

\paragraph*{5. Use shielding selectively.}
Shielding should be applied where it protects mission-critical functions against the residual sub-centimeter and low-centimeter environment, but it should not be mistaken for a system-level solution. The correct use is selective protection of tanks, pressure vessels, avionics, and other failure-critical surfaces, coupled to layout choices that minimize vulnerable area.

\paragraph*{6. Create pricing and incentive mechanisms that reward verified risk reduction.}
A durable control regime requires more than technical standards; it requires financial incentives that approximate the marginal environmental harm created by long-lived objects in congested shells. New missions should face binding performance standards for disposal and passivation, complemented by refundable deposits or performance bonds tied to verified disposal. Legacy remediation should be financed primarily through public procurement or pooled remediation funds. Prepared end-of-life services and high-risk tracking services should be allowed to transition toward commercial subscription or contract models as interfaces and data standards mature.

\subsection{Sequential implementation priorities}

The recommended program is sequential in both control and procurement terms. The first step is compliance internalization for new missions through disposal standards, standardized reporting, and deposit-refund or performance-bond mechanisms. The second step is deployment of prepared end-of-life servicing and higher-precision covariance/tasking services supported by standardized interfaces, operator disclosure, and a stable civil \SSA/\STM baseline. The third step is publicly procured remediation of the legacy stock, in which governments or multilateral consortia purchase verified hazard reduction rather than removal activity alone.

This sequencing matters because each layer relaxes a distinct implementation constraint for the next. Reporting and disposal verification improve auditability and reduce contract uncertainty. Standardized interfaces and a stable civil baseline reduce the execution uncertainty of operator-procured services. Only after those conditions are in place does large-scale remediation of unprepared legacy derelicts become suitable for sustained procurement rather than isolated demonstrations.

\subsection{Limitations and empirical calibration needs}
\label{sec:limitations}

This paper intentionally uses a reduced-order control framework for intervention ranking, procurement logic, and market design rather than a fully calibrated evolutionary environment model. Its equations are therefore intended for intervention ranking, engineering interpretation, and institutional design rather than for stand-alone prediction. The most important unresolved calibration quantities are the fragment-yield mapping $Y_{j\ell\rightarrow s}$, catastrophic-breakup probabilities $P^{\rm cat}_{ij\ell}$, shell-specific flux proxies $\phi_i$, operator threshold distributions $P_c^\star$, and lifecycle cost curves for \PMD, \JCA, and \ADR. A natural next step would be to calibrate Eqs.~(\ref{eq:population_balance})--(\ref{eq:budget_selection}) against a simplified shell-and-size environment dataset and to solve a stylized portfolio problem numerically.

\section{Conclusions}
\label{sec:concl}

Orbital debris in Earth orbit is best understood as an operations-and-stability problem with nonlinear failure dynamics. The present environment is already beyond a regime in which passive self-correction can be assumed: populated \LEO shells continue to grow more congested, fragmentation remains an active source term, and long-horizon scenario analyses show continued debris growth even under no-further-launches assumptions. Three conclusions follow.

First, the problem is already visible in operator workload. Starlink maneuver reporting demonstrates that conjunction management at constellation scale has become an industrial operations function rather than an occasional flight-dynamics activity. Second, the dominant long-horizon hazard is concentrated not in the numerically largest class but in a relatively small stock of inactive high-mass objects in persistent shells. Third, mitigation alone is not enough once the legacy stock is large; the effective control variables are disposal reliability for new missions, encounter-state uncertainty for the high-risk conjunction tail, and the hazard stock of legacy objects.

In programmatic terms, the resulting control stack is sequential rather than simultaneous. The first priority is to drive $p_{\rm disp}$ upward for all new missions in crowded shells. The second is to reduce the operational inflation associated with $C_b$ in the high-risk conjunction tail through targeted sensing, covariance realism, and standardized coordination. The third is to reduce the inherited stock $\Pi_k$ of inactive intact mass through selective \JCA or \ADR. This ordering matters because each layer improves the cost-effectiveness of the next. 

The present framework is comparative rather than predictive. It is intended for intervention ranking, engineering interpretation, and institutional design; it is not a substitute for calibrated full-fidelity evolutionary environment models such as LEGEND. Its specific contribution is the mapping from three reduced-order control quantities---$(1-p_{\rm disp})\tau_k$ for newly launched missions, $\sqrt{\det C_b}$ for the high-risk conjunction tail, and $\Pi_k$ for inactive high-mass legacy objects---to implementation variables that are observable, auditable, and contractible.

Those implementation variables are verified disposal, verified reduction in ambiguous high-risk conjunctions, and verified reduction in residual hazard stock. They should define reporting requirements, procurement specifications, payment clauses, and allocation of responsibility. The corresponding institutional allocation is asymmetric: operators should internalize mission-generated risk through reliable disposal and passivation; governments and multilateral consortia should finance common-good \SSA/\STM and initial remediation of the legacy stock; and prepared end-of-life servicing should be procured directly by operators where interfaces and consent are already defined. The metric of success is not the number of objects removed or the number of remediation missions flown, but verified reduction in time-integrated environmental hazard in the shells and object classes that dominate long-horizon instability.

Taken together, the analysis reduces near-term orbital sustainability to three operational imperatives. First, new missions must cease adding persistent dead mass to congested shells, which means disposal in less than 5 years should be treated as a floor rather than an aspirational target. Second, the operational burden of the current environment should be attacked through targeted reduction of encounter-state uncertainty rather than through pursuit of perfect custody of every object. Third, the long-horizon instability of the environment requires selective intervention on the relatively small stock of inactive high-mass objects that dominates future fragment production. The efficient institutional allocation is therefore asymmetric: operators should internalize mission-generated risk through reliable disposal and passivation; governments and multilateral consortia should finance common-good \SSA/\STM and first-wave remediation of the legacy stock; and prepared end-of-life servicing should evolve into a hybrid commercial market. The metric of success is not the number of objects removed, but verified reduction in time-integrated environmental hazard.

The main result of the paper is therefore comparative. Near-term debris control can be organized around three quantities: $(1-p_{\rm disp})\tau_k$ for newly launched missions, $\sqrt{\det C_b}$ for the high-risk conjunction tail, and $\Pi_k$ for inactive high-mass legacy objects. These variables map directly into implementation metrics---verified disposal, verified reduction in ambiguous high-risk conjunctions, and verified reduction in hazard stock---and therefore into procurement specifications, payment clauses, and allocation of responsibility. That mapping is the paper's central technical contribution.

\section*{Acknowledgments}
The work described here was carried out at the Jet Propulsion Laboratory, California Institute of Technology, Pasadena, California, under a contract with the National Aeronautics and Space Administration. 
 
\bibliographystyle{apsrev4-2}

%

\end{document}